\documentclass[pra,twocolumn,superscriptaddress,nobibnotes,letterpaper]{revtex4}

\usepackage[pdftex]{graphicx}
\usepackage[pdftex]{epsfig}
\usepackage{amsmath}
\usepackage{amssymb}
\usepackage{amsfonts}
\usepackage{color}
\usepackage{wrapfig}
\usepackage{eucal}
\usepackage{hhline}
\usepackage{threeparttable}
\usepackage{supertabular}
\usepackage{multirow}
\usepackage{tabularx}

\newcommand{\opd}[2]{\mbox{$\hat{#1}_{#2}^{\dagger}$}}
\newcommand{\opdagger}[2]{\mbox{$\hat{#1}_{#2}^{\dagger}$}}
\newcommand{\op}[2]{\mbox{$\hat{#1}_{#2}$}}

\def\be{\begin{equation}}
\def\ee{\end{equation}}
\def\bea{\begin{eqnarray}}
\def\eea{\end{eqnarray}}

\newcommand{\avg}[1]{\mbox{$\langle#1\rangle$}}

\newcolumntype{Y}{>{\centering\arraybackslash}X}

\newcommand{\opX}[3]{\mbox{$\hat{#1}^{(#2)}_{#3}$}}

\begin{document}

\pagenumbering{arabic}

\title{Squeezing of light via reflection from a silicon micromechanical resonator}

\author{Amir H. Safavi-Naeini}
\thanks{These authors contributed equally to this work.}
\author{Simon Gr\"oblacher}
\thanks{These authors contributed equally to this work.}
\author{Jeff T. Hill}
\thanks{These authors contributed equally to this work.}
\affiliation{Kavli Nanoscience Institute and Thomas J. Watson, Sr., Laboratory of Applied Physics, California Institute of Technology, Pasadena, CA 91125}
\affiliation{Institute for Quantum Information and Matter, California Institute of Technology, Pasadena, CA 91125}
\author{Jasper Chan}
\affiliation{Kavli Nanoscience Institute and Thomas J. Watson, Sr., Laboratory of Applied Physics, California Institute of Technology, Pasadena, CA 91125}
\author{Markus Aspelmeyer}
\affiliation{Vienna Center for Quantum Science and Technology (VCQ), Faculty of Physics, University of Vienna, A-1090 Wien, Austria}
\author{Oskar Painter}
\email{opainter@caltech.edu}
\homepage{http://copilot.caltech.edu}
\affiliation{Kavli Nanoscience Institute and Thomas J. Watson, Sr., Laboratory of Applied Physics, California Institute of Technology, Pasadena, CA 91125}
\affiliation{Institute for Quantum Information and Matter, California Institute of Technology, Pasadena, CA 91125}
\affiliation{Max Planck Institute for the Science of Light, G\"unther-Scharowsky-Stra\ss{}e 1/Bldg.\ 24, D-91058 Erlangen, Germany}

\date{\today}


\begin{abstract}
We present the measurement of squeezed light generation using an engineered optomechanical system fabricated from a silicon microchip and composed of a micromechanical resonator coupled to a nanophotonic cavity.  Laser light is used to measure the fluctuations in the position of the mechanical resonator at a measurement rate comparable to the free dynamics of the mechanical resonator, and greater than its thermal decoherence rate. By approaching the strong continuous measurement regime we observe, through homodyne detection, non-trivial modifications of the reflected light's vacuum fluctuation spectrum. In spite of the mechanical resonator's highly excited thermal state ($10,000$ phonons), we observe squeezing at the level of $4.5 \pm 0.5\%$ below that of shot-noise over a few MHz bandwidth around the mechanical resonance frequency of $28$~MHz. This squeezing is interpreted as an unambiguous quantum signature of radiation pressure shot-noise.
\end{abstract}

\maketitle


Monitoring a mechanical object's motion, even with the gentle touch of light, fundamentally alters its dynamics. The experimental manifestation of this basic principle of quantum mechanics, its link to the quantum nature of light, and the extension of quantum measurement to the macroscopic realm have all received extensive attention over the last half century~\cite{Braginsky1995,Clerk2010}. The use of squeezed light, with quantum fluctuations below that of the vacuum field, was proposed nearly three decades ago~\cite{Caves1981} as a means for beating the standard quantum limits in precision displacement and force measurements. Conversely, it has also been proposed that a strong continuous measurement of a mirror's position with light, may itself give rise to squeezed light~\cite{Fabre1994,Mancini1994}. In this Letter, we present such a continuous position measurement using an engineered optomechanical system fabricated from a silicon microchip and composed of a micromechanical resonator coupled to a nanophotonic cavity.  Laser light is used to measure the fluctuations in the position of the mechanical resonator at a measurement rate comparable to the free dynamics of the mechanical resonator, and greater than its thermal decoherence rate. By approaching the strong continuous measurement regime we observe, through homodyne detection, non-trivial modifications of the reflected light's vacuum fluctuation spectrum. In spite of the mechanical resonator's highly excited thermal state ($10^4$ phonons), we observe squeezing at the level of $4.5 \pm 0.5\%$ below that of shot-noise over a few MHz bandwidth around the mechanical resonance frequency of $\omega_m/2\pi \approx 28$~MHz. This squeezing is interpreted as an unambiguous quantum signature of radiation pressure shot-noise~\cite{Purdy2013}.

The generation of states of light with fluctuations below that of vacuum has been of great theoretical interest since the 1970s~\cite{Yuen1976,Hollenhorst1979,Caves1981,Walls1983}. Early experimental work demonstrated squeezing of a few percent in a large variety of different nonlinear systems, such as neutral atoms in a cavity~\cite{Slusher1985}, optical fibers~\cite{Shelby1986}, and crystals with bulk optical nonlinearities~\cite{Wu1986,Grangier1987}. This initial research was mainly pursued as a strategy to mitigate the effects of shot-noise given the possibility of improved optical communication~\cite{Yuen1976} and better sensitivity in gravitational wave detectors~\cite{Hollenhorst1979,Caves1981}. In recent years, in addition to being installed in gravitational wave detectors~\cite{Ligo2011}, squeezed light has enhanced metrology in more applied settings~\cite{Taylor2013}.

The vacuum fluctuations arising from the quantum nature of light determine our ability to optically resolve mechanical motion and set limits on the perturbation caused by the act of measurement~\cite{Caves1980b}. A well-suited system to experimentally study quantum measurement is that of cavity-optomechanics, where an optical cavity's resonance frequency can be designed to be sensitive to the position of a mechanical system. By monitoring the phase and intensity of the reflected light from such a cavity, a continuous measurement of mechanical displacement can be made. Systems operating on this simple premise have been realized in a variety of experimental settings, such as in large-scale laser gravitational wave interferometers~\cite{Abbott2009}, microwave circuits with electromechanical elements~\cite{Rocheleau2010,Teufel2011b}, mechanical elements integrated with or comprising Fabry-P\'{e}rot cavities~\cite{Gigan2006,Arcizet2006b,Corbitt2006}, and on-chip nanophotonic cavities sensitive to mechanical deformations~\cite{Eichenfield2009a,Chan2011}. Conceptually, the same basic optomechanical interaction appears when the collective motion of an ultracold gas of atoms shifts the resonance frequency of an optical cavity, and recent gas-phase experiments have shown a variety of optomechanical effects, including squeezed light generation~\cite{Brooks2012}.

The simplest cavity-optomechanical system consists of two modes, an optical and a mechanical resonance, and is parameterized by their respective resonance frequencies ($\omega_\text{o}$, $\omega_\text{m}$) and quality factors ($Q_\text{o}$, $Q_\text{m}$), and the coupling rate of intracavity light to the resonant mechanical motion ($g_0$). The Hamiltonian describing the interaction between light and mechanics is $H_\text{int} = \hbar g_0 \opd{a}{}\op{a}{} \op{x}{}/x_\text{zpf}$, with the amplitude of the mechanical resonator motion being given by $\op{x}{} = x_\text{zpf}\cdot(\opd{b}{} + \op{b}{})$, where $x_\text{zpf} = \sqrt{\hbar/2m_{\mathrm{eff}}\omega_{\mathrm{m}}}$ is the zero-point fluctuation and $m_{\mathrm{eff}}$ the effective motional mass of the resonator. Here $\op{a}{}$ ($\opd{a}{}$) and $\op{b}{}$ ($\opd{b}{}$) are the annihilation (creation) operators of optical and mechanical excitations, respectively. The optical cavity decay rate, $\kappa=\omega_\text{o}/Q_\text{o}$, is the loss rate of photons from the cavity and the rate at which optical vacuum fluctuations, or shot-noise, is coupled into the optical resonance~\cite{Gardiner1985}. Similarly, the mechanical damping rate $\gamma_\text{i}=\omega_\text{m}/Q_\text{m}$ is the rate at which thermal bath fluctuations couple into the mechanical system. In all experimental realizations of optomechanics to date, including that presented here, the optomechanical coupling rate $g_0$ has been much smaller than the cavity decay rate $\kappa$. As such, without a strong coherent drive, the interaction of the vacuum fluctuations with the mechanics is negligible. 

Under the effect of a coherent laser drive, the cavity is populated with a mean intracavity photon number $\avg{n_\text{c}}$, and one considers the optical fluctuations about the classical steady-state, $\op{a}{} \rightarrow \sqrt{\avg{n_\text{c}}} + \op{a}{}$. This modifies the optomechanical interaction resulting in a linear coupling between the fluctuations of the intracavity optical field ($\op{X}{0} = \op{a}{} + \opd{a}{}$) and the position fluctuations of the mechanical system $\op{x}{}$:\ $H_\text{int} = \hbar G \op{X}{0}\op{x}{}/x_\text{zpf}$. The parametric linear coupling occurs at an effective interaction rate of $G \equiv \sqrt{\avg{n_\text{c}}}g_0$. Through this interaction, the intensity fluctuations of the vacuum field $\opX{X}{\text{in}}{\theta=0}(t)$ entering the cavity impart a force on the mechanical system,
\bea
\op{F}{\text{BA}}(t) = \frac{\hbar\cdot\sqrt{\Gamma_\text{meas}}}{x_\text{zpf}}\opX{X}{\text{in}}{\theta=0}(t)\label{eqn:FBA},
\eea
which is the radiation pressure shot-noise (RPSN) force. The mechanical motion in turn is recorded in the phase of the light leaving the cavity,
\bea
\opX{X}{\text{out}}{\theta}(t) = -\opX{X}{\text{in}}{\theta}(t)  - 2 \frac{\sqrt{\Gamma_\text{meas}}}{x_\text{zpf}} \op{x}{}(t) \cdot\sin(\theta),\label{eqn:Xout_dynamic}
\eea
where $\theta$ is the quadrature angle of the detected optical field, with $\theta=0$ ($\pi/2$) referring to the intensity (phase) quadrature. The latter can be interpreted as the informational aspect of measurement, with the optical cavity playing the role of the position detector (measuring observable $\op{x}{}$ at a rate $\Gamma_\text{meas} \equiv 4 G^2/\kappa$), while the former is the quantum measurement back-action imposing a fluctuating noise force onto the mechanical system~\cite{Clerk2010}. In addition to the back-action noise, thermal fluctuations from the bath also drive the mechanical motion, with their magnitudes becoming comparable as $\Gamma_\text{meas}$ approaches the thermalization rate $\Gamma_\text{thermal}(\omega) \equiv \gamma_\text{i} \bar{n}(\omega)$. The bath occupation number $\bar{n}(\omega)$ is equivalent to its high temperature limit $k_\text{B} T_b / \hbar \omega$ in these measurements. 

The auto-correlation of the optical output field has contributions due to the auto-correlation of the optical input field and the mechanical fluctuations, as well as their cross-correlations. Due to the back-action force on the mechanical system, the motion has non-zero correlations with the RPSN ($\avg{\op{x}{}(t) \op{F}{\text{BA}}(t')}$), which can be negative for some values of $\theta$. Under certain conditions this can lead to the squeezing of the optical output field. The spectral density of a balanced homodyne measurement of the output field, normalized to shot-noise, is given in the quasi-static limit for frequencies $\omega \ll \omega_{\text{m}}$ (see the Appendix):
\bea
\bar{S}^\mathrm{out}_{II}(\omega) &=& 1 + 4(\Gamma_\text{meas}/\omega_\text{m}) \sin(2\theta)\nonumber\\&&
 +4\frac{\Gamma_\text{meas}}{\omega_\text{m}}\frac{\bar{n}(\omega)}{Q_m}(1+\cos(2\theta)).
\label{eqn:approx_thermal_SII}
\eea
Measurement of the spectrum below shot-noise (squeezing) is therefore attributable to cross-correlations of measurement back-action noise due the optical vacuum fluctuations and position fluctuations of the mechanical resonator.  

\begin{figure}[t!]
\begin{center}
\includegraphics[width=1.\columnwidth]{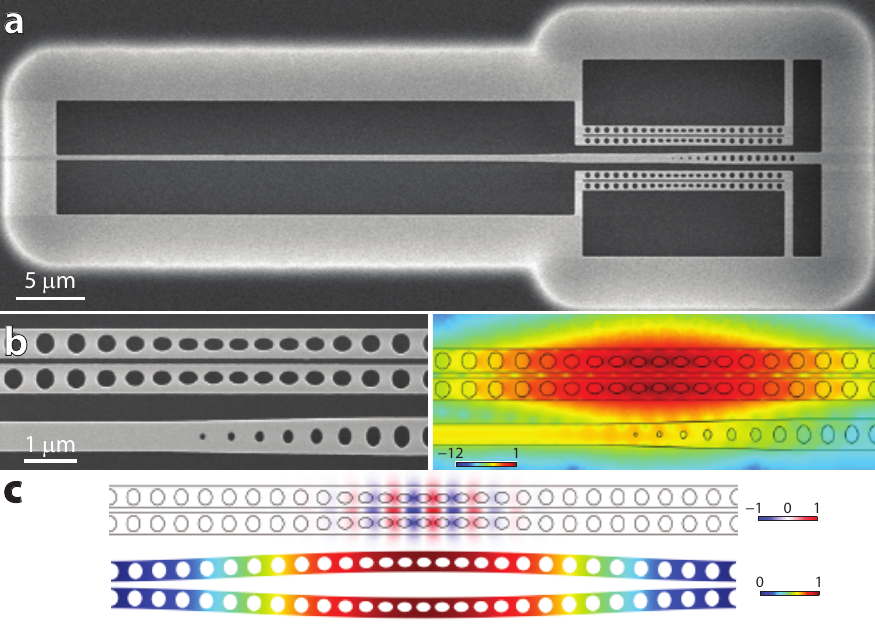}
\caption{\textbf{Optomechanical device.} \textbf{a}, Scanning electron microscope image of a waveguide-coupled zipper optomechanical cavity. The waveguide width is adiabatically tapered along its length and terminated with a photonic crystal mirror next to the cavity. The tapering of the waveguide allows for efficient input/output coupling while the photonic crystal termination makes the coupling to the cavity single-sided. Two zipper cavities are coupled above and below the waveguide, each with a slightly different optical resonance frequency allowing them to be separately addressed. \textbf{b}, (left) Close-up of the coupling region between one of the cavities and the waveguide. (right) Finite element method (FEM) simulation of the cavity field leaking into the waveguide (log scale). Note that the field does not leak into the mirror region of the waveguide. \textbf{c}, (top) FEM simulation showing the in-plane electrical field of the fundamental optical cavity mode. (bottom) FEM simulation of the displacement of the fundamental in-plane differential mode of the structure with frequency $\omega_\text{m}/2\pi=28$~MHz. The mechanical motion, modifying the gap between the beams, shifts the optical cavity frequency leading to optomechanical coupling.}
\label{fig:1}
\end{center}
\end{figure}

\begin{figure*}[ht!]
\begin{center}
\includegraphics[width=2\columnwidth]{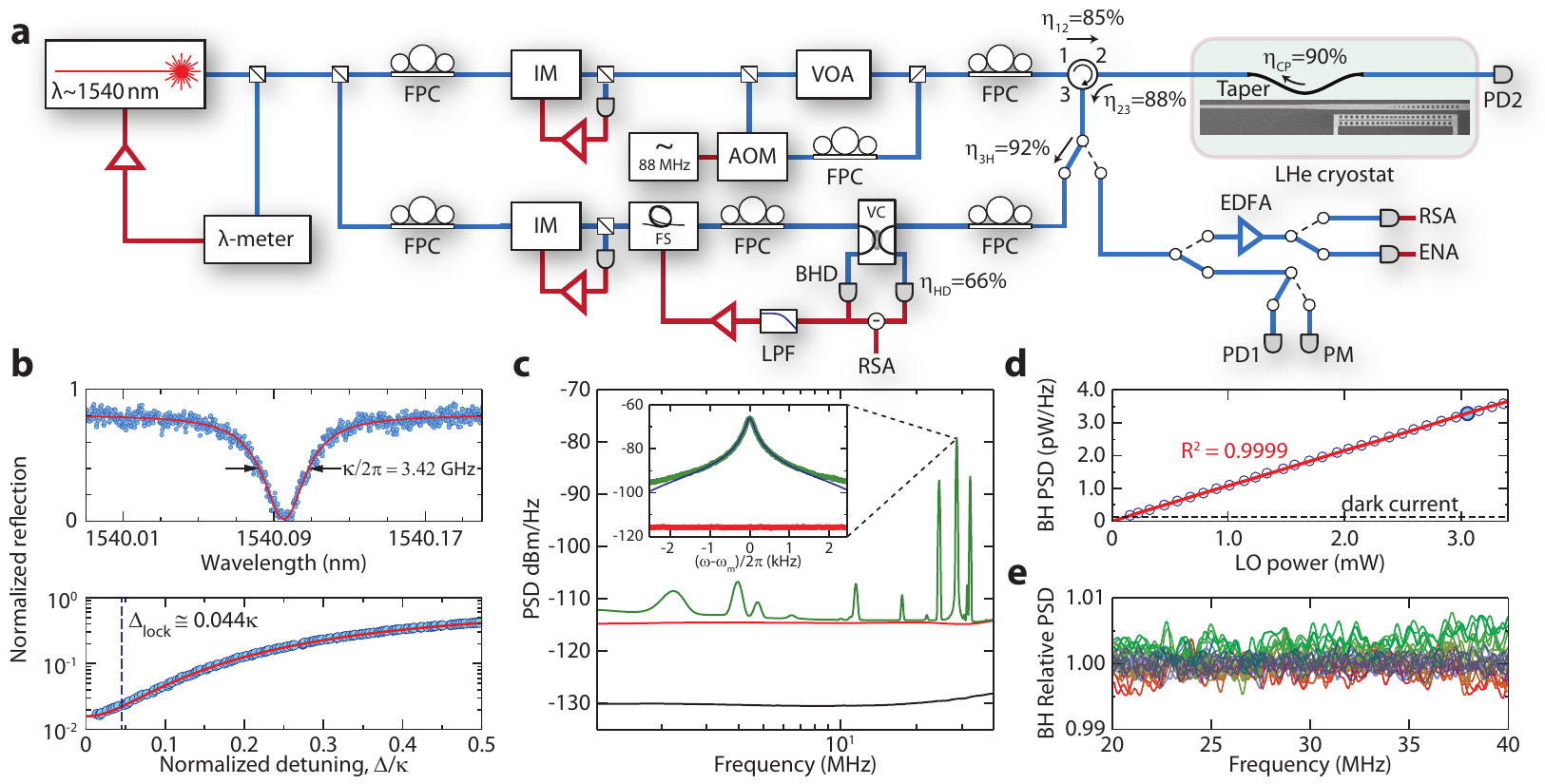}
\caption{\textbf{Experimental setup and device characterization.} \textbf{a} A tunable external-cavity diode laser, actively locked to a wavemeter ($\lambda$-meter), is used to generate a strong local oscillator (LO) and the measurement signal. A portion of each field is split off, detected, and used to individually stabilize the intensity in each arm using intensity modulators (IM). Fiber polarization controllers (FPC) adjust the polarization of the LO and signal. A variable optical attenuator (VOA) is used to set the signal power and an acousto-optic modulator (AOM) is used to generate a tone of known amplitude for calibration (see Appendix). The signal beam is fed into a fiber taper that is mounted inside a $^4$He continuous flow cryostat and is coupled to the zipper cavity. The reflected signal beam from the cavity is separated using a circulator, and is switched between one of three detection paths: a power meter (PM) for power calibration, a photodetector (PD1) for optical spectroscopy of the cavity, and an erbium-doped fiber amplifier (EDFA) for spectroscopy of the coupled cavity-optomechanical system on a real-time spectrum analyzer (RSA) or network analyzer (ENA). For the measurement of squeezing, the reflected cavity signal is sent to a fourth path containing a variable coupler (VC), where it is recombined with the strong local oscillator and detected on a balanced homodyne detector (BHD). The relative phase between the LO and the signal is set using a fiber stretcher (FS). \textbf{b} (top) Reflected signal from the optical cavity at low optical power ($n_c\approx 10$, linewidth $\kappa/2\pi=3.42$~GHz).  (bottom) High-power ($n_c=790$) reflected signal, showing the cavity-laser detuning (dashed line) locked to during squeezing measurements. \textbf{c} Homodyne noise PSD of the reflected signal showing the transduced thermal Brownian motion of the zipper cavity at $T_\text{b}=16$~K (green curve; $n_c=80$). The red curve is the shot noise level and the black curve is the detector's dark noise. The inset shows a zoom-in of the fundamental in-plane differential mechanical mode of the zipper cavity (linewidth $\gamma_\text{i}/2\pi=172$~Hz). \textbf{d} Mean value of the PSD of the balanced homodyne detector as a function of the LO power (signal blocked). The filled data point indicates the LO power used in the squeezing measurements. The red and dashed black curves correspond to a linear fit to the data and the detector dark current level, respectively. \textbf{e} Noise PSD as a function of $\theta_{\text{lock}}$ with the signal detuned far off-resonance at $\Delta/\kappa \approx 30$ (shades green to red), referenced to the noise level with the signal blocked (blue).}
\label{fig:2}
\end{center}
\end{figure*}

The primary technical hurdle to observing such squeezing, as in many quantum measurements, is the strong coupling of a preferred detection channel (the optical probe) simultaneous with the minimization of unwanted environmental perturbations of the mechanical system. As indicated in Eq.~(\ref{eqn:approx_thermal_SII}), fluctuations from the thermal bath limit squeezing of the optical probe field to a regime in which $\bar{n}(\omega) < Q_\text{m}$. This requirement is equivalent to having a $Q$-frequency product, $Q_{\text{m}} \omega_\text{m}  > k_\text{B} T_\text{b} / \hbar$. Equation~(\ref{eqn:approx_thermal_SII}) also indicates that squeezing becomes appreciable only as the measurement rate of the mechanics by the light field approaches the mechanical frequency. In some respects, both of these challenges have been overcome in recent optomechanics work~\cite{Teufel2011b,Chan2011,Verhagen2012}, whereby thermal bath coupling has been made smaller than the optically-induced cooling of the mechanical resonator. However, significant squeezing over an appreciable spectral bandwidth requires not only a large cooperativity between light and mechanics, as represented by $C=\Gamma_{\text{meas}}/\gamma_i$ and realized in cooling experiments, but the more stringent requirement that the effective measurement back-action force be comparable to all forces acting on the mechanics, including the elastic restoring force of the mechanical structure. A more comprehensive model (see Appendix), including the resonant response of the mechanical system to RPSN, shows that the strict requirements of the quasi-static model are somewhat relaxed, and that the squeezing scales approximately as $\Gamma_{\text{meas}}/\delta$, where $\delta$ is the effective bandwidth of squeezing around the mechanical resonance frequency.





In order to meet the requirements of strong measurement and efficient detection, we designed a zipper-style optomechanical cavity~\cite{Eichenfield2009a} with a novel integrated waveguide coupler fabricated from the $220$~nm thick silicon device layer of a silicon-on-insulator microchip (see Fig.~\ref{fig:1}a). The in-plane differential motion of the two beams at a fundamental frequency of $\omega_{\mathrm{m}}/2\pi = 28~\mathrm{MHz}$ strongly modulates the co-localized fundamental optical resonance of the cavity with a theoretical vacuum coupling rate of $g_0/2\pi = 1~\mathrm{MHz}$.  As shown in Fig.~\ref{fig:1}{b}, we use a silicon waveguide with a high reflectivity photonic crystal end-mirror to efficiently excite and collect light from the optical cavity. Coupling light from the silicon waveguide to a single-mode optical fiber is performed using an optical fiber taper and a combination of adiabatic mode coupling and transformation (see Fig.~\ref{fig:1}{b}).



The experimental setup used to characterize the zipper cavity system and measure the optomechanical squeezing of light is shown in Fig.~\ref{fig:2}a. The silicon sample is placed in a continuous flow $^4$He cryostat with a cold finger temperature of $10$~K.  A signal laser beam is used to probe the optomechanical system and measure the mechanical motion of the zipper cavity. A wavelength scan of the reflected signal from the cavity is plotted in Fig.~\ref{fig:2}c, showing an optical resonance with a linewidth $\kappa/2\pi=3.42$~GHz at a wavelength of $\lambda_c=1540$~nm. Inefficiencies in the collection and detection of light correspond to additional uncorrelated shot-noise in the signal and can reduce the squeezing to undetectable levels. For the device studied here, the cavity coupling efficiency, corresponding to the percentage of photons sent into the cavity which are reflected, is determined to be $\eta_k=0.54$. The fiber-to-chip coupling efficiency is measured at $\eta_\text{CP} = 0.90$. A homodyne detection scheme~\cite{Yuen1983} allows for high efficiency detection of arbitrary quadratures of the optical signal field. Characterization and optimization of the efficiency of the entire optical signal path and homodyne detection system (see Appendix for details) results in an overall setup efficiency of $\eta_\text{setup} = 0.48$, corresponding to a total signal detection efficiency of $\eta_\text{tot} = \eta_\text{setup} \eta_\kappa = 0.26$.  

Figure~\ref{fig:2}c shows the noise spectrum of the thermal motion of the mechanical resonator obtained by positioning the laser frequency near the cavity resonance and tuning the relative local oscillator (LO) phase of the homodyne detector, $\theta_{\text{lock}}$, to measure the quadrature of the reflected signal in which mechanical motion is imprinted (roughly the phase quadrature for near-resonance probing). The mechanical spectrum is seen to contain the in-plane differential mode of interest at $\omega_m/2\pi=28$~MHz, as well as several other more weakly coupled mechanical resonances of the nanobeams and coupling waveguide (the in-plane differential mode peak appears reduced relative to the other modes in this plot due to the limited resolution bandwidth of the measurement).  A high-resolution, narrowband spectrum of the in-plane differential mode is displayed as an inset to Fig.~\ref{fig:2}c, and shows a linewidth of $\gamma_i/2\pi=172$~Hz, corresponding to a mechanical $Q$-factor of $Q_\text{m}=1.66 \times 10^{5}$.  The vacuum coupling rate of the in-plane differential mode, measured from the detuning dependence of the optical spring shift and damping (see Appendix), is determined to be $g_0/2\pi=750$~kHz, in good correspondence with theory. From the calibration of the noise power under the Lorentzian in Fig.~\ref{fig:2}c, the in-plane differential mode is found to thermalize (at low optical probe power) to a temperature of $T_{\mathrm{b}} \sim 16$~K, corresponding to a phonon occupancy of $\avg{n} \sim 1.2\times 10^{4}$.  This yields a ratio, $Q_{\text{m}} \hbar \omega_\text{m}/k_\text{B} T_\text{b} \approx 13$, well within the regime where coherent motion and squeezing are possible.     



\begin{figure*}[t]
\begin{center}
\includegraphics[width=2\columnwidth]{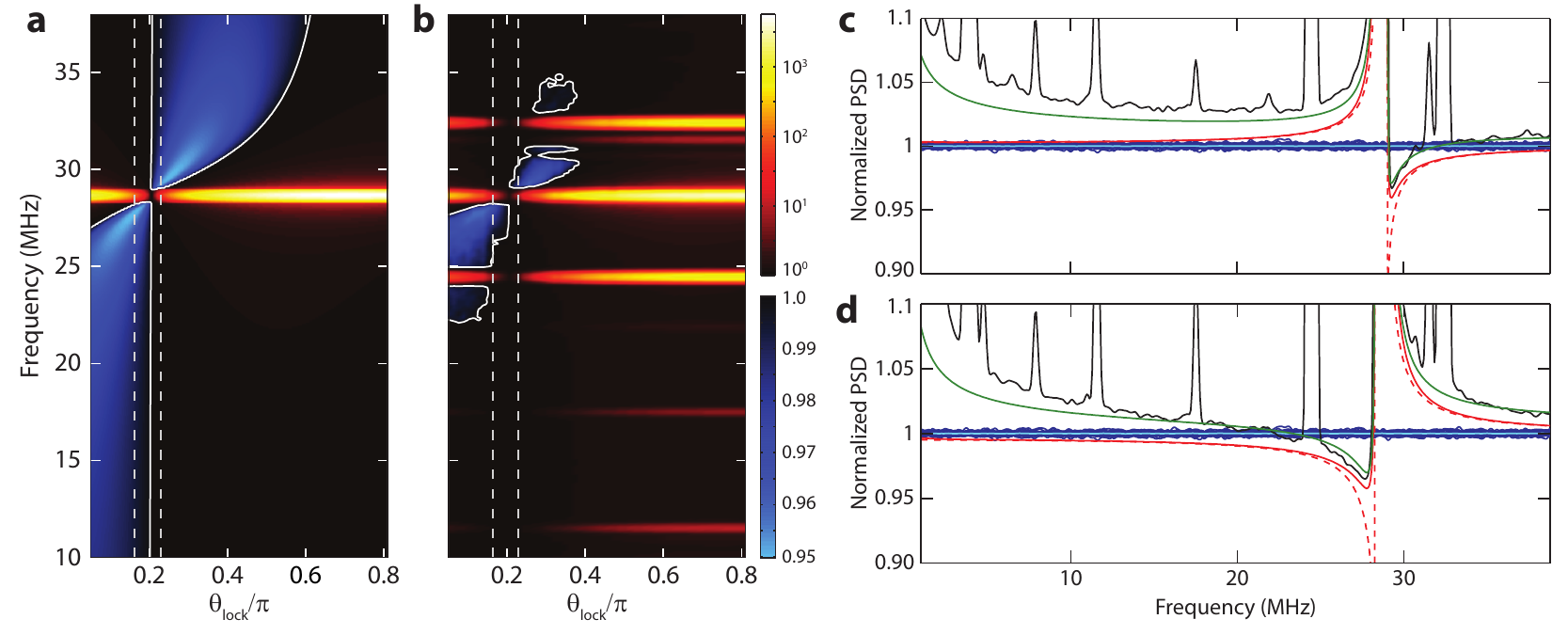}
\caption{\textbf{Optomechanical squeezing of light.} \textbf{a}, Theoretical model. Density plot of the predicted reflected signal noise PSD, as measured on a balanced homodyne detector and normalized to shot-noise, for a simplified model of the optomechanical system (see Appendix). Areas below shot-noise are shown in blue shades on a linear scale. Areas with noise above shot-noise are shown in orange shades on a log-scale. The solid white line is a contour delineating noise above and below shot-noise. \textbf{b}, Experimental data. Density plot of the measured reflected signal noise PSD for $n_c=790$ normalized to the measured shot-noise level. \textbf{c}, Slice of the measured density plot in \textbf{b} taken at $\theta_{\text{lock}}/\pi=0.23$.  \textbf{d}, Slice of the measured density plot in \textbf{b} taken at $\theta_{\text{lock}}/\pi=0.16$.  In \textbf{c} and \textbf{d}, the black curve corresponds to the measured data slice extracted from \textbf{b}.  The dark blue traces are several measurements of the shot-noise level (average shown in light blue).  Also indicated is a model of the squeezing in the absence of thermal noise (red dashed curve), the same model with thermal noise included (solid red curve), and a full noise model including additional phenomenological noise sources (solid green curve).}
\label{fig:3}
\end{center} 
\end{figure*}

In order to systematically and accurately study the noise properties of the reflected optical signal from the cavity we make a series of measurements to characterize our laser and detection setup. Figure~\ref{fig:2}d shows the measured noise power spectral density (PSD) of the balanced homodyne detector (dark current subtracted) for $\omega \approx \omega_m$ as a function of LO power (signal blocked), indicating a linear dependence on power and negligible added noise above shot-noise. In the measured squeezing data to follow, a LO power of $3$~mW is used. Calibration of the laser intensity and frequency noise over the frequency range of interest ($\omega/2\pi=1$--$40$~MHz) is measured by direct photodetection of the laser, pre- and post-transmission through a fiber Mach-Zehnder interferometer with a known frequency response. The laser intensity noise is measured to be shot-noise dominated over this frequency range, while the laser frequency noise is measured to be roughly flat at a level of $S_{\omega \omega} \sim 5\times 10^3$~rad$^2 \cdot$Hz. Laser phase noise on the signal beam can be converted to intensity noise by reflection from the dispersive cavity or due to frequency dependent components in the optical train, and can add to the detected noise floor. Due to the broad linewidth of the zipper cavity resonance, the effects of the laser phase noise are found to be negligible. The full suite of noise and calibration measurements performed are described in detail in the Appendix.

\begin{figure}[t]
\begin{center}
\includegraphics[width=1\columnwidth]{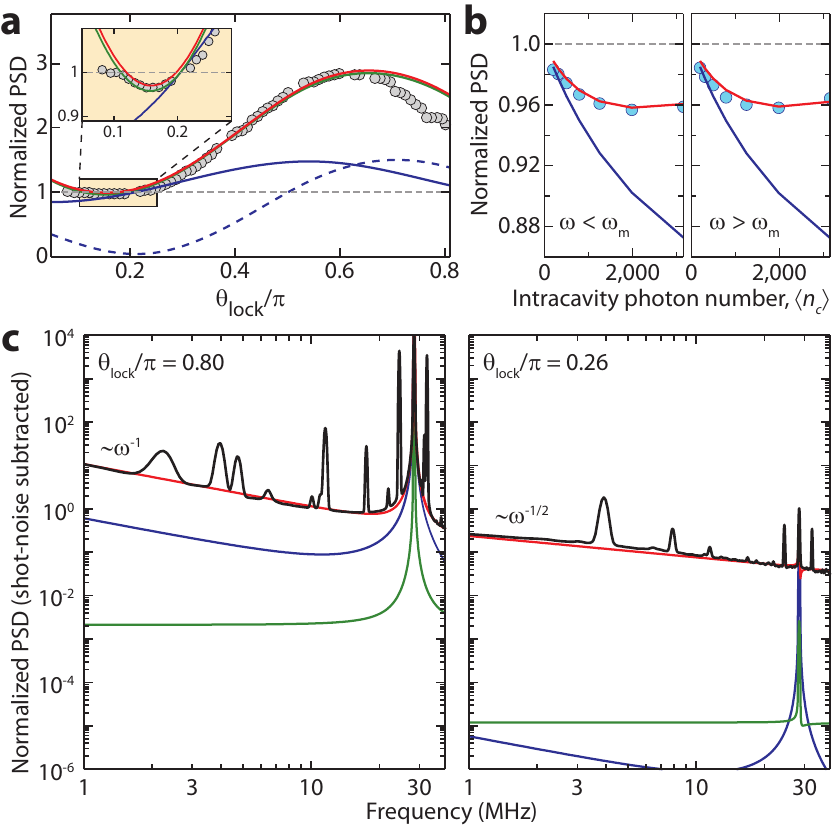}
\caption{\textbf{Spectral and power dependence of noise.} \textbf{a}, Measured (filled circles) balanced homodyne noise power of the reflected signal at $\omega/2\pi=27.9$~MHz versus quadrature angle ($\Delta_{\text{lock}}/\kappa = 0.044$ and $n_c=790$). The red curve corresponds to the full noise model. The solid blue curve is for a model including the response of the mechanical mode in the absence of thermal noise, i.e., driven by RPSN only (the dashed blue curve shows the thermal noise component). \textbf{b}, Measured (filled circles) minimum noise PSD normalized to shot-noise versus $n_c$. The left plot is the maximum squeezing for $\omega < \omega_m$ and the right for $\omega > \omega_m$. Also shown is the single-mode noise model (blue curve) and the full noise model (red curve). \textbf{c}, Balanced homodyne noise PSD of the reflected cavity signal for $\Delta_{\text{lock}}/\kappa=0.052$ and $n_c=3,153$. Left (right) plot shows phase quadrature corresponding to maximum (minimum) transduction of mechanical motion. The black curve is the measured data with the shot-noise level subtracted.  Also shown are modeled laser phase noise (green curve), the single-mode noise model (blue curve), and the full noise model (red curve).}
\label{fig:4}
\end{center} 
\end{figure}

Measurement of the noise in the reflected optical signal from the cavity as a function of quadrature angle, frequency, and signal power is presented in Figs.~\ref{fig:3} and \ref{fig:4}. These measurements are performed for laser light on resonance with the optical cavity and for input signal powers varying from $252$~nW to $3.99$~$\mu$W in steps of $2$~dB, with the maximum signal power corresponding to an average intra-cavity photon number of $\avg{n_\text{c}} = 3,153$. Positioning of the laser at the appropriate cavity detuning for each signal power is performed by scanning the wavelength across the cavity resonance while monitoring the reflection, and then stepping the laser frequency towards the cavity from the red side until the reflection matches the level that corresponds to a detuning of $\Delta_{\text{lock}}/\kappa \approx 0.04$ (see Fig.~\ref{fig:2}b).  The laser is locked to this frequency using a wavemeter with a frequency resolution of $\pm 0.0015\kappa$.  Drift of the optical cavity resonance over the timescale of a single noise spectrum measurement (minutes) is found to be negligible. An estimate of the variance of $\Delta_{\text{lock}}$ is determined from the dependence of the transduction of the mechanical motion on the quadrature phase, indicating that from one lock to another $\Delta_{\text{lock}}/\kappa = 0.044 \pm 0.006$.

In Fig.~\ref{fig:3} we plot the theoretically predicted and measured noise PSD versus quadrature angle for a signal power corresponding to $\avg{n_c}=790$~photons. Each quadrature spectrum is the average of 150 traces taken over 20 seconds, and after every other spectrum, the signal arm is blocked and the shot-noise PSD is measured. The shot-noise level, which represents the noise of the electromagnetic vacuum on the signal arm, is used to normalize the spectra. We find at certain quadrature angles, and for frequencies a few MHz around the mechanical resonance frequency, that the light reflected from the zipper cavity shows a noise PSD below that of vacuum. The density plot of the theoretically predicted noise PSD (Fig.~\ref{fig:3}a) shows the expected wideband squeezing due to the strong optomechanical coupling in these devices, as well as a change in the phase angle where squeezing is observed at below and above the mechanical frequency. This change is due to the change in sign of the mechanical susceptibility and the corresponding change in phase of the mechanical response to RPSN. The measured noise PSD density plot (Fig.~\ref{fig:3}b) shows the presence of several other mechanical noise peaks and a reduced squeezing bandwidth, yet the overall phase- and frequency-dependent characteristics of the squeezing around the strongly-coupled in-plane mechanical mode are clearly present. In particular, Figs.~\ref{fig:3}c and d show two slices of the noise PSD density plot which show the region of squeezing change from being below to above the mechanical resonance frequency. 


In Fig.~\ref{fig:4}a we show the measured noise PSD (grey circles) as a function of quadrature angle for a frequency slice at $\omega/2\pi=27.9$~MHz of the data shown in Fig.~\ref{fig:3}b. The measured squeezing (anti-squeezing) is seen to be smaller (larger) than expected from a model of the optomechanical cavity without thermal noise. We also plot in Fig.~\ref{fig:4}b the maximum measured and modeled squeezing as a function of signal power. The simple theory predicts a squeezing level (blue curve) which monotonically increases with signal power, whereas the measured maximum squeezing saturates at a level of $4.5 \pm 0.5\%$ below the shot-noise at an intracavity power $\avg{n_\text{c}} = 1,984$ photons. In order to understand the noise processes that limit the bandwidth and magnitude of the measured squeezing, we plot in Fig.~\ref{fig:4}c the noise PSD (shot-noise subtracted) for phase quadratures that maximize (left plot) and minimize (right plot) the transduction of the mechanical mode peak at $\omega_m/2\pi=28$~MHz. Along with the measured data (black curve), we also plot the estimated noise due to phase noise of the signal laser (green curve) and that for a single mechanical mode (blue curve) assuming a thermal bath temperature of $T_b =16$~K and a frequency-independent damping rate. The single-mode noise model greatly underestimates the background noise level, especially in the quadrature minimizing transduced motion in which we measure a $\omega^{-1/2}$ (as opposed to $\omega^{-1}$) frequency dependence to the low frequency noise.  As described in the Appendix, the additional measured noise is thought to arise from a combination of the thermal noise tails of higher frequency mechanical modes and fluctuations in the optical cavity damping rate, along with a small amount of heating due to optical absorption in the silicon cavity. The red curves in each of the plots in Fig.~\ref{fig:4} show the full noise model incorporating these added phenomenological terms.


Somewhat surprisingly, these measurements show that by reflecting light off a thin-film mechanical resonator undergoing large amplitude thermal motion, light that is in certain respects quieter than vacuum can be obtained. This is found to result from radiation-pressure fluctuations being strongly imprinted on and modified by the motion of the mechanical resonator, thus demonstrating a fundamentally quantum aspect of displacement measurement. The devices in this work utilize lithographic patterning at the nanoscale to transform silicon into a material with an effective quantum optical nonlinearity at an engineerable optical wavelength.  The modest level of squeezing realized in this work is predominantly limited by thermal noise, but also by the efficiency with which the reflected light can be collected by external optics.  The effects of thermal noise may be substantially reduced by working with materials of higher intrinsic $Q$-frequency product, such as diamond and silicon carbide~\cite{Ayazi2011}.  Given the microchip form of the devices studied here, and the potential for device integration, it is interesting to consider whether new squeezed light applications might arise. For example, squeezed light generated by one device could be directly sent into another device for use as an optical probe.  Such an on-chip squeezer and detector could be used as a quantum-enhanced micro-mechanical displacement and force sensor~\cite{Hoff2013}.  More generally, we expect future experiments with feedback and strong measurement of the dynamics of a mechanical system to be within reach.  In addition, using quantum light as an input we expect to be able to generate entangled states of mechanics and light with similar devices.



The authors would like to thank K.\ Hammerer and A.\ A.\ Clerk for valuable discussions. This work was supported by the DARPA/MTO ORCHID program through a grant from AFOSR, Institute for Quantum Information and Matter, an NSF Physics Frontiers Center with support of the Gordon and Betty Moore Foundation, and the Kavli Nanoscience Institute at Caltech. ASN and JC gratefully acknowledge support from NSERC. SG acknowledges support from the European Commission through a Marie Curie fellowship.

\def\urlprefix{}
\def\url#1{}

{
\newcommand{\nocontentsline}[3]{}
\renewcommand{\addcontentsline}[2][]{\nocontentsline#1{#2}}

}

\newpage

\makeatletter 
\def\tagform@#1{\maketag@@@{(A\ignorespaces#1\unskip\@@italiccorr)}}
\makeatother

\makeatletter
\makeatletter \renewcommand{\fnum@figure}
{\figurename~A\thefigure}
\makeatother

\renewcommand{\figurename}{Figure}

\clearpage
\onecolumngrid

\begin{center}

\Large{
\textbf{Appendix}}

\vspace{0.2in}

\end{center}

\twocolumngrid

\setcounter{figure}{0}%
\setcounter{equation}{0}%
\setcounter{section}{0}
\setcounter{tocdepth}{3}
\tableofcontents

\section{Theory}\label{sec:theory_squeezing}

Optomechanical systems can be described theoretically with the Hamiltonian (see main text)
\bea
H = \hbar \omega_\text{o} \opdagger{a}{} \op{a}{} + \hbar \omega_{\text{m}0} \opdagger{b}{} \op{b}{} + \hbar g_0 \opdagger{a}{}\op{a}{}(\opdagger{b}{} + \op{b}{}),
\eea
where $\op{a}{}$ and $\op{b}{}$ are the annihilation operators for photons and phonons in the system, respectively. Generally, the  system is driven by intense laser radiation at a frequency $\omega_\text{L}$, making it convenient to work in an interaction frame where $\omega_\text{o}$ is replaced by $\Delta$ in the above Hamiltonian with $\Delta = \omega_\text{o} - \omega_\text{L}$. To quantum mechanically describe the dissipation and noise from the environment, we use the quantum-optical Langevin differential equations (QLEs)~\cite{Gardiner1985,Collett1984,Gardiner2004},
\bea
\dot{\op{a}{}}(t) &=& -\left(i\Delta + \frac{\kappa}{2}\right) \op{a}{} - ig_0 \op{a}{}(\opdagger{b}{}+\op{b}{}) \nonumber\\&&-\sqrt{\kappa_\text{e}} \op{a}{\mathrm{in}}(t) - \sqrt{\kappa_\text{i}}\op{a}{\mathrm{in,i}}(t), \nonumber\\
\dot{\op{b}{}}(t) &=& -\left(i\omega_{\text{m}0} + \frac{\gamma_\text{i}}{2}\right) \op{b}{} - ig_0 \opdagger{a}{} \op{a}{} - \sqrt{\gamma_\text{i}}\op{b}{\mathrm{in}}(t),\nonumber
\eea
which account for coupling to the bath with dissipation rates $\kappa_\text{i}$, $\kappa_\text{e}$, and $\gamma_\text{i}$ for the intrinsic cavity energy decay rate, optical losses to the waveguide coupler, and total mechanical losses, respectively. The total optical losses are $\kappa =\kappa_\text{e}+\kappa_\text{i}$. These loss rates are necessarily accompanied by random fluctuating inputs $\op{a}{\mathrm{in}}(t)$, $\op{a}{\mathrm{in,i}}(t)$, and $\op{b}{\mathrm{in}}(t)$, for optical vacuum noise coming from the coupler, optical vacuum noise coming from other optical loss channels, and mechanical noise (including thermal). 

The study of squeezing is a study of noise propagation in the system of interest and as such, a detailed understanding of the noise properties is required. The equations above are derived by making certain assumptions about the noise, and are generally true for the case of an optical cavity, where thermal noise is not present, and where we are interested only in a bandwidth of roughly $10^8$ smaller than the optical frequency (0 -- 40~MHz bandwidth of a 200~THz resonator). For the mechanical system, where we operate at very large thermal bath occupancies ($\gg10^3$) and are interested in the broadband properties of noise sources (0 -- 40~MHz for a 30~MHz resonator), a more detailed understanding of the bath is required, and will be presented in the section on thermal noise.

At this point, we linearize the equations assuming a strong coherent drive field $\alpha_0$, and displace the annihilation operator for the photons by making the transformation $\op{a}{} \rightarrow \alpha_0 + \op{a}{}$. This approximation, which neglects terms of order $\op{a}{}^2$ is valid for systems such as ours where $g_0 \ll \kappa$, i.e.\ the \textit{vacuum weak coupling} regime. We are then left with a parametrically enhanced coupling rate  $G = g_0|\alpha_0|$. Using the relations given in the mathematical definitions section (\ref{sec:math_defn}) of this document,
we write the solution to the QLEs in the Fourier domain as
\bea
\op{a}{}(\omega) &=&  \frac{-\sqrt{\kappa_\text{e}} \op{a}{\mathrm{in}}(\omega) - \sqrt{\kappa_\text{i}}\op{a}{\mathrm{in,i}}(\omega) - iG(\op{b}{}(\omega) + \opdagger{b}{}(\omega))  }{i(\Delta-\omega) + \kappa/2}\label{eqn:opt_fluct},\nonumber \\
\op{b}{}(\omega) &=&  \frac{-\sqrt{\gamma_\text{i}}\op{b}{\mathrm{in}}(\omega)}{i(\omega_{\text{m}0}-\omega) + \gamma_\text{i}/2} -  \frac{i G(\op{a}{}(\omega) + \opdagger{a}{}(\omega))}{i(\omega_{\text{m}0}-\omega) + \gamma_\text{i}/2}.\label{eqn:mech_fluct}
\eea

Finally we note that by manipulation of these equations, the mechanical motion can be expressed as a (renormalized) response to the environmental noise and the optical vacuum fluctuations incident on the optical cavity through the optomechanical coupling

\begin{flalign}
\op{b}{}(\omega) &=  \frac{-\sqrt{\gamma_\text{i}}\op{b}{\mathrm{in}}(\omega)}{i(\omega_\text{m}-\omega) + \gamma/2}\nonumber&\\
&+ \frac{iG}{i(\Delta -\omega) + \kappa/2} \frac{\sqrt{\kappa_\text{e}} \op{a}{\mathrm{in}}(\omega) + \sqrt{\kappa_\text{i}} \op{a}{\mathrm{in,i}}(\omega)}{i(\omega_\text{m}-\omega) + \gamma/2}\nonumber&\\
&+ \frac{iG}{-i(\Delta + \omega) + \kappa/2} \frac{\sqrt{\kappa_\text{e}} \opdagger{a}{\mathrm{in}}(\omega) + \sqrt{\kappa_\text{i}} \opdagger{a}{\mathrm{in,i}}(\omega)}{i(\omega_\text{m}-\omega) + \gamma/2} \label{eqn:b_inputs}.&
\end{flalign}

The renormalized mechanical frequency and loss rate are $\omega_\text{m}=\omega_{\text{m}0}+\delta\omega_\text{m}$, and $\gamma = \gamma_\text{i} + \gamma_\mathrm{OM}$, respectively, with
\begin{flalign}
\delta\omega_\text{m} &= |G|^2 \mathrm{Im}\left[ \frac{1}{i(\Delta-\omega_\text{m})+\kappa/2} - \frac{1}{-i(\Delta+\omega_\text{m})+\kappa/2} \right]\label{eq_spring},\\
\gamma_{\mathrm{OM}} &= 2|G|^2 \mathrm{Re}\left[ \frac{1}{i(\Delta-\omega_\text{m})+\kappa/2} - \frac{1}{-i(\Delta+\omega_\text{m})+\kappa/2} \right]\label{eq_damp}.
\end{flalign}

It is convenient to define here what we mean by a quadrature, as it is the observable of the light field that our measurement device (the balanced homodyne detector (BHD) setup) is sensitive to:
\bea
\opX{X}{j}{\theta}= \op{a}{j} e^{-i \theta}+ \opdagger{a}{j}e^{i \theta}.
\eea
We are interested in the properties of $\opX{X}{\mathrm{out}}{\theta}$ for various quadrature angles $\theta$, given the influence of the mechanical system.

The measurement of the field provides us with a record $\op{I}{}(t) = \opX{X}{\mathrm{out}}{\theta}(t)$ for a certain $\theta$. We use a spectrum analyzer to perform Fourier analysis on this signal and obtain a symmetrized classical power spectral density (PSD) $\bar{S}_{II}(\omega)$, as defined in the mathematical appendix (section \ref{sec:math_defn}).

For a vacuum field such as the input field, the measured quadrature $\opX{X}{\mathrm{vac}}{\theta}(t)$ will have a power spectral density  
\bea
\bar{S}^\mathrm{vac}_{II}(\omega) = 1.
\eea
This is the shot-noise level which is due to the quantum fluctuations of the electromagnetic field. Mathematically, it arises from the correlator $\avg{\op{a}{\mathrm{vac}}(\omega)\opdagger{a}{\mathrm{vac}}(\omega^\prime)} = \delta(\omega+\omega^\prime)$, with all other correlators $\avg{\opdagger{a}{\mathrm{vac}}(\omega)\op{a}{\mathrm{vac}}(\omega^\prime)}$, $\avg{\opdagger{a}{\mathrm{vac}}(\omega)\opdagger{a}{\mathrm{vac}}(\omega^\prime)}$, $\avg{\op{a}{\mathrm{vac}}(\omega)\op{a}{\mathrm{vac}}(\omega^\prime)}$, arising in the expression $\avg{\opdagger{I}{}(\omega)\op{I}{}(\omega^\prime)}$ equal to zero.

\subsection{Approximate quasi-static theory}\label{ss:quasi-static}
\begin{figure*}[ht!]
\begin{center}
\includegraphics[width=2\columnwidth]{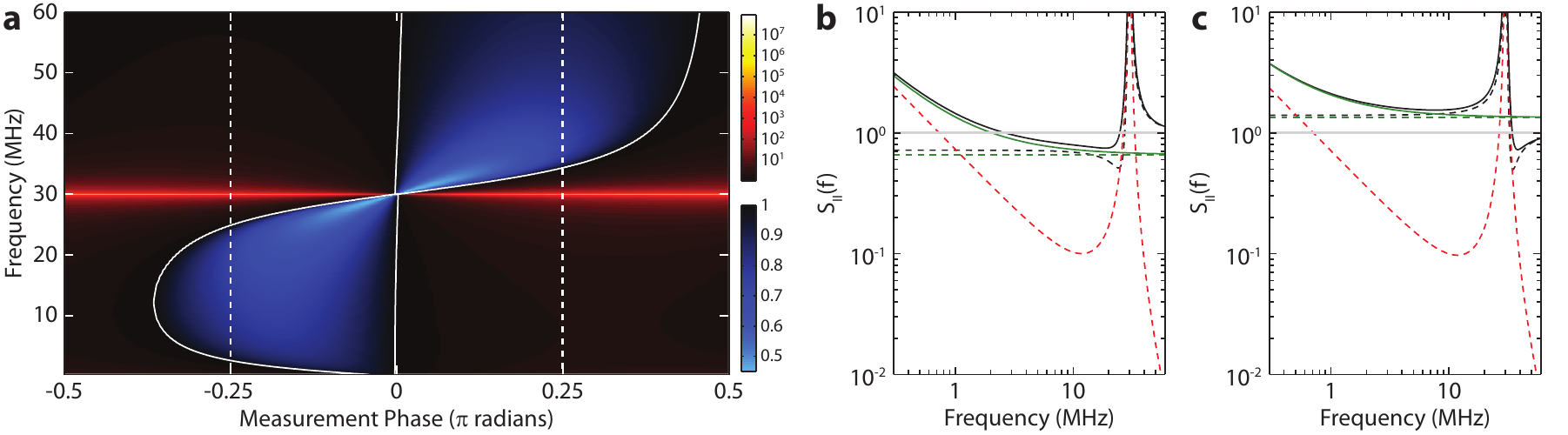}
\caption{\textbf{Squeezing theory.} \textbf{a}, Density plot of the predicted squeezing $\bar{S}^\mathrm{out}_{II}(\omega)$ vs.\ phase angle and frequency, normalized to the shot-noise. The mechanical mode can clearly be seen at $\omega_{\mathrm{m}}/2\pi = 30$~MHz. The solid white lines outline the region where the power spectral density falls below 1 (the shot-noise level) indicating the presence of squeezing for that phase and frequency. The dashed white lines at $\theta = -\pi/4$ and $\theta = +\pi/4$ correspond to regions where squeezing can be obtained below and above the mechanical frequency, respectively, and the components of the noise model for these phases is shown in detail in figures~\textbf{b} and \textbf{c}. In these figures the spectra are again normalized to the shot-noise level plotted as a grey line. The simple squeezing model without thermal noise (Eq.~\eqref{eqn:approx_bare_SII}) is represented by the dashed green line and the simple model with thermal noise (Eq.~\eqref{eqn:approx_thermal_SII}) is the solid green line. The solid black line is the full squeezing model $\bar{S}^\mathrm{out}_{II}(\omega)$ corresponding to \textbf{a} with the constituent components: the contribution from the optical vacuum fluctuations ($\bar{S}^\mathrm{out}_{II,a}(\omega)$; Eq.~\eqref{eqn:SIIaoptvac}) represented by the dashed black line and the thermal noise ($\bar{S}^\mathrm{out}_{II,b}(\omega)$; Eq.~\eqref{eqn:SIIbthermal}) represented by the dashed red line.}
\label{fig:squeeze_theory}
\end{center}
\end{figure*}

In this section we present a simplified derivation of how squeezing is obtained in the studied optomechanical system to elucidate the important system parameters and their role in squeezing. We make a few approximations to simplify the derivation:
\begin{enumerate}
\item $\Delta = 0$: The laser is tuned exactly to the optical cavity frequency.
\item  $\kappa_\text{e} = \kappa$: Perfect coupling. 
\item $\kappa \gg \omega_\text{m}$:  Bad cavity limit.
\item $\omega \ll \omega_\text{m}$: We are only interested in the quasi-static response, so the resonant response of the mechanical resonator does not play a role.
\end{enumerate}

Under these assumptions, equations (\ref{eqn:mech_fluct}) and (\ref{eqn:b_inputs}) can be written as (using the relation for the optical output field $\op{a}{\mathrm{out}}(\omega)=\op{a}{\mathrm{in}}(\omega)  + \sqrt{\kappa}\op{a}{}(\omega))$:
\bea
i\omega_\text{m} \op{b}{}(\omega) &=&  -\sqrt{\gamma_\text{i}}\op{b}{\mathrm{in}}(\omega)
+ \frac{2iG}{\sqrt{\kappa}}(\op{a}{\mathrm{in}}(\omega) + \opdagger{a}{\mathrm{in}}(\omega)),\nonumber\\
\op{a}{\mathrm{out}}(\omega) &=&-\op{a}{\mathrm{in}}(\omega)
- \frac{2iG}{\sqrt{\kappa}}(\op{b}{}(\omega) + \opdagger{b}{}(\omega)).\label{eqn:aoutsimple}
\eea
The first equation shows the response of the mechanical resonator subsystem to the thermal bath fluctuations ($\op{b}{\mathrm{in}}(\omega)$) and the optical vacuum noise from the measurement back-action. We define $\Gamma_\text{meas} \equiv 4|G|^2/\kappa$, and interpret it as the measurement rate~\cite{Clerk2010}, such that the factor appearing in front of the optical vacuum noise operators is $\sqrt{\Gamma_\text{meas}}$.
This rate also appears in the second equation for the output field, in front of the normalized position operator $\op{x}{}/x_\text{zpf} = \op{b}{}(\omega) + \opdagger{b}{}(\omega)$, which is the observable that is being measured.

Note, from the expression for $\op{a}{\mathrm{out}}(\omega)$ it follows, that since the position is a real observable with an imaginary prefactor, the effects we consider depend strongly on the quadrature being probed, i.e.\ the real part of the expression, $\opX{X}{\mathrm{out}}{\theta=0}$, will not be affected by the optomechanical coupling.

At this point we can easily calculate the properties of the detected spectrum $\bar{S}^\mathrm{out}_{II}(\omega)$, by writing $\op{a}{\mathrm{out}}$ in terms of $\op{a}{\mathrm{in}}$ and $\op{b}{\mathrm{in}}$ for which the correlators are known:
\bea
\op{a}{\mathrm{out}}(\omega) &=&-\op{a}{\mathrm{in}}(\omega)
- 2i\frac{\Gamma_\text{meas}}{\omega_\text{m}} (\op{a}{\mathrm{in}}(\omega) + \opdagger{a}{\mathrm{in}}(\omega))\nonumber\\&&
+ \frac{\sqrt{\gamma_\text{i} \Gamma_\text{meas}}}{\omega_\text{m}}(\op{b}{\mathrm{in}}(\omega) - \opdagger{b}{\mathrm{in}}(\omega)).\label{eqn:approx_aout}
\eea
Ignoring thermal noise for the moment ($\gamma_\text{i} = 0$), and dropping terms of order $(\Gamma_\text{meas}/\omega_\text{m})^2$ (assuming $\Gamma_\text{meas} \ll \omega_\text{m}$) we arrive at:
\bea
\bar{S}^\mathrm{out}_{II}(\omega) &=& \int_{-\infty}^{\infty} d\omega^\prime~ \avg{\opX{X}{\text{out}}{\theta}(\omega)\opX{X}{\text{out}}{\theta}(\omega^\prime)} \nonumber\\
&=& 1 + 4(\Gamma_\text{meas}/\omega_\text{m}) \sin(2\theta).\label{eqn:approx_bare_SII}
\eea
Note that for certain values of $\theta$, the detected spectral density can be smaller than what one would expect for a vacuum field. For $\theta = - \pi/4$, we achieve the maximum squeezing with a noise floor of $1-4(\Gamma_\text{meas}/\omega_\text{m})$ which strongly dependends on the ratio $\Gamma_\text{meas}/\omega_\text{m}$.

To understand the effect of thermal noise, we assume the form of the correlator to be $\avg{\op{b}{\mathrm{in}}(\omega)\opdagger{b}{\mathrm{in}}(\omega^\prime)} = (\bar{n}(\omega) + 1) \delta(\omega+\omega^\prime)$, $\avg{\opdagger{b}{\mathrm{in}}(\omega)\op{b}{\mathrm{in}}(\omega^\prime)} = \bar{n}(\omega) \delta(\omega+\omega^\prime)$, $\avg{\opdagger{b}{\mathrm{in}}(\omega)\opdagger{b}{\mathrm{in}}(\omega^\prime)} = 0$, and $\avg{\op{b}{\mathrm{in}}(\omega)\op{b}{\mathrm{in}}(\omega^\prime)} = 0$ (these expressions are discussed in section~\ref{ss:phenom_thermal_noise_model}). Then a calculation similar to the one leading to equation (\ref{eqn:approx_bare_SII}) gives
\bea
\label{eqn:approx_thermal_SII}
\bar{S}^\mathrm{out}_{II}(\omega) &=& 1 + 4(\Gamma_\text{meas}/\omega_\text{m}) \sin(2\theta)\nonumber\\&&
 +4\frac{\Gamma_\text{meas}}{\omega_\text{m}}\frac{\bar{n}(\omega)}{Q_m}(1+\cos(2\theta)),
\eea
where we have assumed $\bar{n}(\omega)$, the bath occupation at frequency $\omega$, to be much larger than unity. At $\theta = -\pi/4$, we have
\bea
\bar{S}^\mathrm{out}_{II}(\omega) =1 - 4(\Gamma_\text{meas}/\omega_\text{m}) (1 - \bar{n}(\omega)/{Q_m}).
\eea
In this model, there is no squeezing at $\omega$ if $\bar{n}(\omega)  > Q_m$. This means that for squeezing to be present, coherent evolution of the mechanical system must be possible, i.e.\ the rate at which phonons enter the mechanical system from the bath ($\gamma_\text{i} \bar{n}$) must be smaller than the mechanical frequency $\omega_\text{m}$. In conclusion, the important requirements to achieve squeezing are to make $\Gamma_\text{meas}$ comparable to $\omega_\text{m}$, and to reduce the thermal occupancy or increase the mechanical quality factor to achieve $\bar{n}(\omega)  < Q_m$.

\subsection{The effect of dynamics and correlation between RPSN and position}\label{ss:dynamic_squeezing_derivation}

As a next step, we take into account the dynamics of the mechanical resonator while keeping the approximations of the bad-cavity limit ($\kappa\gg\omega_\text{m}$) and on-resonant probing ($\Delta = 0$). In addition to further clarifying some of the observed features, this treatement, as presented in the main text, elucidates the role of correlations between the mechanical system's position and the back-action force.

The response of the mechanical system to a force is captured by its susceptibility:
\bea
\chi_\text{m} (\omega) = \frac{1}{m(\omega_\text{m}^2 - \omega^2 - i \gamma_\text{i} \omega_\text{m})}. \label{eqn:chi_m}
\eea
The form of the damping considered here is the strongly sub-ohmic structural damping which is observed in our measurements~\cite{Saulson1990,Gillespie1993} (cf.\ Section~\ref{ss:phenom_thermal_noise_model}). The mechanical system responds to random noise forces $F_\text{T}(t)$ from the thermal bath (which we treated in the last section and neglect here), and to the quantum back-action from the cavity $F_\text{BA}(t)$.

The back-action force for the resonant case can simply be found by linearizing the expression for the radiation pressure force $\op{F}{\text{RP}}(t) = -\hbar g \opdagger{a}{}\op{a}{}/x_\text{zpf}$. We find the force imparted on the mechanics due to the shot-noise of the cavity field to be
\bea
\op{F}{\text{BA}}(t) = \frac{\hbar\cdot\sqrt{\Gamma_\text{meas}}}{x_\text{zpf}}\opX{X}{\text{in}}{\theta=0}(t)\label{eqn:FBA}
\eea
for the case of resonant driving. The fluctuations imparted on the mechanics are from the intensity quadrature of the light ($\theta = 0$). Using equation~(\ref{eqn:aoutsimple}), we can write the output field quadrature as:
\bea
\opX{X}{\text{out}}{\theta}(t) = -\opX{X}{\text{in}}{\theta}(t)  - 2 \frac{\sqrt{\Gamma_\text{meas}}}{x_\text{zpf}} \op{x}{}(t) \cdot\sin(\theta).\label{eqn:Xout_dynamic}
\eea
We note here that the mechanical position fluctuations are primarily imprinted on the phase quadrature of the output light, with $\theta = \pm\pi/2$. The intensity quadrature is unmodified ($\opX{X}{\text{out}}{\theta=0}(t) = -\opX{X}{\text{in}}{\theta=0}(t)$) since changes in the cavity frequency are not transduced as changes in intensity when the laser is resonant with the cavity.

The output of the homodyne detector normalized to the shot-noise level is found by taking the auto-correlation of eqn.~\eqref{eqn:Xout_dynamic}. The correlations between radiation pressure shot-noise and the mechanical motion are important in this calculation~\cite{Heidmann1997,Verlot2009,Borkje2010,Safavi-Naeini2012,Khalili2012,Jayich2012,Safavi-Naeini2013a} and must be taken into account.  In the time-domain we find the auto-correlation to be:
\begin{widetext}
\bea
\avg{\opX{X}{\text{out}}{\theta}(t)\opX{X}{\text{out}}{\theta}(t')} &=& \delta(t - t')  + 4 \Gamma_\text{meas} \sin^2(\theta) \frac{\avg{\op{x}{}(t) \op{x}{}(t')}}{x^2_\text{zpf}} \nonumber\\&&+ 2 \hbar^{-1} \sin(\theta) \cos(\theta) \avg{ \op{F}{\text{BA}}(t) \op{x}{}(t') + \op{x}{}(t) \op{F}{\text{BA}}(t')}. \label{eqn:XoutXout}
\eea
\end{widetext}
The $\cos(\theta)$ in the last term comes from the general expression for a quadrature $\opX{X}{\text{in}}{\theta}(t) = \opX{X}{\text{in}}{\theta=0}(t) \cos(\theta) + \opX{X}{\text{in}}{\theta=\pi/2}(t) \sin(\theta)$, and equation~\eqref{eqn:FBA}. The key components of equation~\eqref{eqn:XoutXout} are the shot-noise level, the thermal noise, and the cross-correlation between the back-action noise force and mechanical position fluctuations. It is only the latter which can give rise to squeezing, by reducing the fluctuation level below shot-noise. This squeezing can be calculated spectrally:
\bea
S_\text{sq}(\omega) &=& \hbar^{-1} \sin(2\theta) \times \int_{-\infty}^\infty \text{d}\tau [\avg{ \op{F}{\text{BA}}(t) \op{x}{}(t-\tau)}\nonumber\\&& + \avg{\op{x}{}(t) \op{F}{\text{BA}}(t-\tau)}] e^{i\omega \tau} \nonumber\\
&=& 2 \hbar \sin(2\theta)   \frac{{\Gamma_\text{meas}}}{x^2_\text{zpf}} \chi_\text{m}(\omega)\nonumber\\
&=& 4 m \omega_\text{m} \sin(2\theta)   {\Gamma_\text{meas}} \chi_\text{m}(\omega).
\eea
At the DC or quasi-static limit ($\omega \rightarrow 0$) the susceptibility $\chi_\text{m} \rightarrow 1/m\omega_\text{m}^2$ can be used and we reobtain the results from section~\ref{ss:quasi-static} (cf.\ equation~\eqref{eqn:approx_bare_SII}). We see that for $\theta < 0$,  squeezing is obtained in this limit. At frequencies larger than $\omega_\text{m}$, $\chi_\text{m}(\omega)$ changes sign, and we expect to see squeezing at quadrature angles $\theta > 0$. Additionally, since $\chi_\text{m}(\omega)$ becomes larger around the mechanical frequency, we expect the maximum squeezing to be enhanced. More specifically, at a detuning $\delta = \omega_m-\omega$ from the mechanical resonance, we expect the parameter characterizing the squeezing to be proportional to $\Gamma_\text{meas}/\delta$. All of these features are seen in Fig.~\ref{fig:squeeze_theory}.

It is important to note here that in the absence of other nonlinearities in the system, any reduction of the noise below the vacuum fluctuations can only be caused by the correlations between the RPSN and the position fluctuations of the system. This makes the problem of proving the correlations between RPSN and mechanical motion equivalent to the problem of proving that the reflected light from the optomechanical cavity has been squeezed. 

Conceptually this form of probing the RPSN is similar to that carried out by Safavi-Naeini et al.~\cite{Safavi-Naeini2012,Safavi-Naeini2013a} and analyzed by Khalili et al.~\cite{Khalili2012}.  It also shares features with the cross-correlation measurements proposed by Heidmann et al.~\cite{Heidmann1997}, and B\o{}rkje et al.~\cite{Borkje2010}, and recent experiments by Purdy et al.~\cite{Purdy2013}. The distinguishing feature of this type of measurement is that the quantum correlations between the fluctuations of the position and the electromagnetic vacuum manifest themselves as squeezed light.

\subsection{General derivation of squeezing}\label{ss:detailed_squeezing_derivation}

Among the approximations made in  section~\ref{ss:quasi-static}, the quasi-static approximation is the least correct. In fact, in our experiments, the most observable squeezing occurs with $\omega$ close to $\omega_\text{m}$ and even slightly larger than $\omega_\text{m}$, so $\omega \ll \omega_\text{m}$ is not valid. Near the mechanical frequency, resonant enhancement of the optical vacuum fluctuations by the mechanical resonator causes squeezing greater than that predicted in the quasi-static regime to be possible.

Here we show the results of a derivation that does not rely on most of the assumptions used in the approximate model. Of the assumptions in the previous section, the only simplification we keep here is to assume perfect coupling $\kappa_\text{e} = \kappa$. The effect of imperfect coupling can be taken into account trivially and is explained after this section (see~\ref{ss:imperfect_coupling}).

By substitution of equation (\ref{eqn:b_inputs}) into the equation for $\op{a}{}(\omega)$ (\ref{eqn:mech_fluct}), we arrive at:
\bea
\sqrt{\kappa}\op{a}{}(\omega) &=& A_1(\omega) \op{a}{\text{in}}(\omega) + A_2(\omega) \opdagger{a}{\text{in}}(\omega) \nonumber\\&& +B_1(\omega) \op{b}{\text{in}}(\omega) +B_2(\omega) \opdagger{b}{\text{in}}(\omega),
\eea
with
\begin{widetext}
\bea
A_1(\omega) &=& \frac{\kappa}{i(\Delta-\omega)+\kappa/2}\times\nonumber\\&&
\left[\frac{|G|^2}{i(\Delta-\omega)+\kappa/2}\frac{1}{i(\omega_\text{m}-\omega)+\gamma/2}  - \frac{|G|^2}{i(\Delta-\omega)+\kappa/2}\frac{1}{-i(\omega_\text{m}+\omega)+\gamma/2} - 1\right]\\
A_2(\omega) &=& \frac{\kappa}{i(\Delta-\omega)+\kappa/2}\times\nonumber\\&&
\left[\frac{|G|^2}{-i(\Delta+\omega)+\kappa/2}\frac{1}{i(\omega_\text{m}-\omega)+\gamma/2}  - \frac{|G|^2}{-i(\Delta+\omega)+\kappa/2}\frac{1}{-i(\omega_\text{m}+\omega)+\gamma/2}\right]\\
B_1(\omega) &=& \frac{\sqrt{\kappa\gamma_\text{i}}}{i(\Delta-\omega)+\kappa/2} 
\left[\frac{iG}{i(\omega_\text{m}-\omega)+\gamma/2}\right] \\
B_2(\omega) &=& \frac{\sqrt{\kappa\gamma_\text{i}}}{i(\Delta-\omega)+\kappa/2} 
\left[\frac{iG}{-i(\omega_\text{m}+\omega)+\gamma/2}\right] 
\eea
\end{widetext}
The expressions give us the output field in terms of the input fields, since 
\bea
\op{a}{\mathrm{out}}(\omega)&=&\op{a}{\mathrm{in}}(\omega)  + \sqrt{\kappa}\op{a}{}(\omega))\nonumber\\
&=&(1+A_1(\omega)) \op{a}{\text{in}}(\omega) + A_2(\omega) \opdagger{a}{\text{in}}(\omega) \nonumber\\&& +B_1(\omega) \op{b}{\text{in}}(\omega) +B_2(\omega) \opdagger{b}{\text{in}}(\omega).
\eea
We can calculate $\bar{S}^\mathrm{out}_{II}(\omega)$ from this expression, which we split into two parts, one only due to the optical vacuum fluctuations, and the other containing the contribution from thermal noise: $\bar{S}^\mathrm{out}_{II}(\omega) = \bar{S}^\mathrm{out}_{II,a}(\omega) + \bar{S}^\mathrm{out}_{II,b}(\omega)$.
\begin{widetext}
\bea
\bar{S}^\mathrm{out}_{II,a}(\omega) &=& |A_2(-\omega)|^2 + |1+A_1(\omega)|^2 + 2\mathrm{Re}\{e^{-2i\theta}(1+A_1(\omega))A_2(-\omega) \}\label{eqn:SIIaoptvac} \\ 
\bar{S}^\mathrm{out}_{II,b}(\omega) &=& |B_1(\omega)|^2 (\bar{n}(\omega)+1) + |B_1(-\omega)|^2 \bar{n}(\omega)\nonumber\\&&
+|B_2(-\omega)|^2 (\bar{n}(\omega)+1) + |B_2(\omega)|^2 \bar{n}(\omega)\nonumber\\&&
+2\mathrm{Re}\{e^{-2i\theta}B_1(\omega)B_2(-\omega) \} (\bar{n}(\omega)+1)  +2\mathrm{Re}\{e^{-2i\theta}B_1(-\omega)B_2(\omega) \} \bar{n}(\omega) \label{eqn:SIIbthermal}
\eea
\end{widetext}

\subsubsection{The effect of imperfect optical coupling and inefficient detection}\label{ss:imperfect_coupling}

At every juncture in an experiment where the optical transmission efficiency is less than unity ($\eta<1$), an equivalent optical circuit can be defined involving an $\eta:(1-\eta)$ beam splitter with the output being $\eta$ times the input and $(1-\eta)$ times the vacuum. Therefore the effect of optical losses and coupling inefficiencies on the detected spectra can be calculated by replacing the measured field quadrature with:
\bea
\opX{X}{\text{det}}{\theta} = \sqrt{\eta} \opX{X}{\text{out}}{\theta} + \sqrt{1-\eta}\opX{X}{\text{vac}}{\theta}
\eea
This source of vacuum noise is completely unrelated to the cavity output, and there are no cross-correlation terms, so the detected current spectral density will be given by
\bea
\bar{S}^\mathrm{det}_{II}(\omega)  = \eta \bar{S}^\mathrm{out}_{II}(\omega) + (1-\eta)\bar{S}^\mathrm{vac}_{II}(\omega),
\eea
where $\bar{S}^\mathrm{vac}_{II}(\omega) = 1$ is the shot-noise.

Measurement inefficiencies take two forms, one is due to ineffeciencies in the detection, while the second is because of excess electronic noise or ``dark current'' present due to the circuitry of the detector and amplifier. This excess noise can also be thought of as a detection inefficiency by considering the amount of optical shot-noise inserted into the signal which would produce it. Since the dark-current is measured with no optical input, and the real shot-noise level increases linearly with the local oscillator (LO) power, this inefficiency is power dependent and can be minimized for large LO powers. In our case, the dark current was found to be 10.4~dB below the detected shot-noise. The total detector efficiency was measured to be $\eta_{\textrm{HD}}=66\%$.

\section{Experiment}


\begin{figure*}[ht!]
\begin{center}
\includegraphics[width=2\columnwidth]{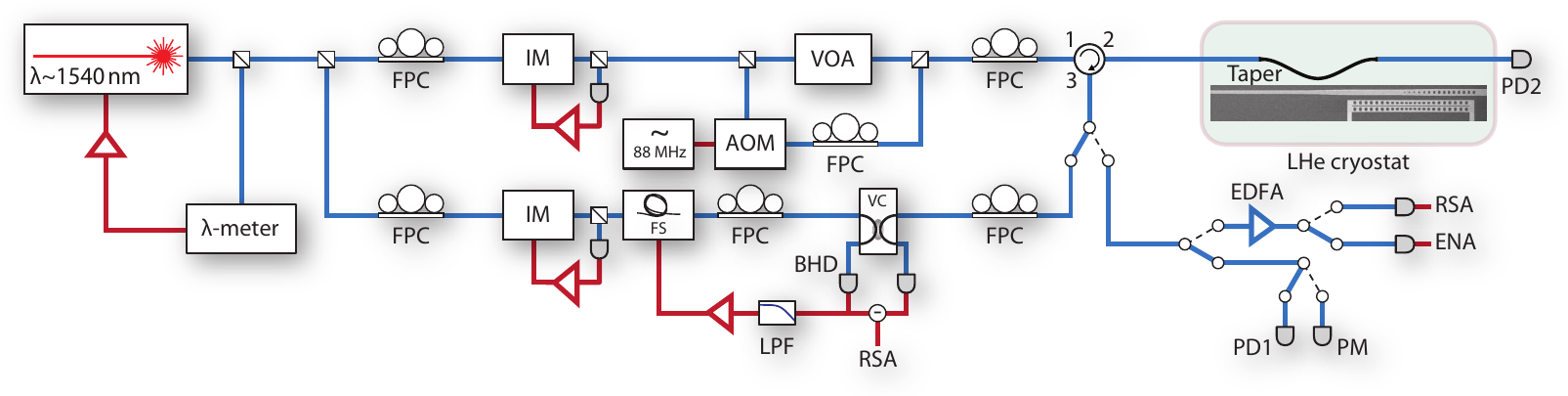}
\caption{\textbf{Experimental setup.} A detailed description of the experimental setup can be found in the main text.}
\label{fig:setup}
\end{center}
\end{figure*}


\subsection{Measurement of losses}
In order to estimate the total squeezing expected in our setup we carefully characterize all losses in our system. Some of these losses are static (e.g.\ circulator losses) while others can vary from experiment to experiment (e.g.\ coupling efficiency of the fiber taper to the waveguide). In figure~\ref{fig:setup} typical losses are shown as efficiencies ($\eta$) for various parts of the experiment. The efficiency of sending light from port 1 to 2 of our optical circulator is $\eta_{12}=85\%$, and $\eta_{23}=88\%$ for port 2 to 3. In addition, the efficiency from port 3 of the circulator to the homodyne detector is $\eta_{\textrm{3H}}=92\%$. All these losses are fixed and do not change over time as the components are optically spliced together. Measuring the coupling efficiency of the fiber taper to the waveguide is done every time a new data set is taken. This is accomplished by switching the light that is reflected from the waveguide to a power meter and comparing the reflected power to a known input power with the laser tuned off-resonance from the optical mode (off-resonance the device acts as a near-perfect mirror). Typical achieved efficiencies are around $\eta_{\textrm{CP}}=90\%$. The efficiency of the homodyne detection strongly depends on the alignment of the polarization between the local oscillator and the signal, as well as by how much the power in the LO overcomes the electronic noise floor of the detector. To determine this efficiency we use an acousto-optic modulator (AOM) inserted in our setup before the circulator in the signal path. The AOM shifts the frequency of the light creating a tone 88~MHz away from the signal with a fixed, known amplitude, and identical polarization to the signal (we directly measure the power of this tone with a power meter). This tone can now be used to determine the total homodyne efficiency by measuring its power on the spectrum analyzer, taking the other losses into account. Our typical homodyne efficiency is $\eta_{\textrm{HD}}=66\%$ resulting in a total setup efficiency (detection efficiency of optical signal photons in the on-chip waveguide) of roughly $\eta_{\textrm{Setup}} = \eta_{\textrm{CP}}\cdot \eta_{23}\cdot \eta_{\textrm{3H}}\cdot \eta_{\textrm{HD}} \approx48\%$.

\begin{figure}[ht!]
\begin{center}
\includegraphics[width=1.\columnwidth]{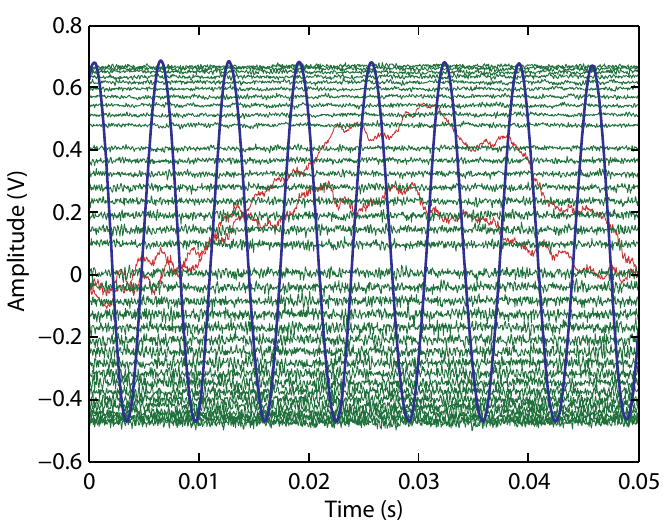}
\caption{\textbf{Phase information.} The blue trace shows the interference signal of the local oscillator and the signal on the homodyne detector when their relative phase is scanned using a fiber stretcher in the local oscillator (LO) arm. The voltage reading here is proportional to $\cos(\theta - \phi)$ where $\theta$ is the phase difference between the LO and input to the cavity, and $\phi$ is the phase imparted by reflection off the cavity. This interference signal is used to actively stabilize the relative phase to different set points (green traces). Occasionally the lock fails, as shown by the red traces, and any associated data is discarded. The range in which the phase can be stably locked is slightly smaller than $-\pi/2$ to $\pi/2$ due to the turning points in the sinusoidal interference curve.}
\label{fig:phases}
\end{center}
\end{figure}

\subsection{Data collection procedure}
Careful calibration of our data is crucial in understanding all noise sources and potential drifts over time in our setup. The losses in our setup are determined before we make a new data run as described in the previous section. We then proceed to record an optical trace of the cavity resonance by switching the light to a photodetector (PD1 in figure~\ref{fig:setup}) and scanning the laser wavelength. This trace provides the information to lock the laser to a fixed detuning (typically $0.04\cdot\kappa$ red of the cavity resonance), which is accomplished using a simple software lock and feedback from the wavemeter (with a resolution of roughly $0.003\cdot\kappa$) and is described in more detail in the subsection below.  As a next step the optical signal is switched to the homodyne detector and the relative phase between the signal and the local oscillator is scanned using the fiber stretcher in the LO arm. The resulting interference is shown in figure~\ref{fig:phases} as the blue trace. The interference signal is  used to lock the relative phase between the signal and LO using a Toptica DigiLock 110. The green traces show the properly locked signal, while the red traces are phase set points where the lock failed requiring the associated data to be discarded. We then record the spectra of the homodyne signal and for every trace taken we also save a spectrum of the shot-noise by switching the signal arm away from the homodyne detector and only measure vacuum input to the signal arm of our detector.  We re-lock the laser with respect to the cavity every other data point to counteract drift. This procedure is repeated for several different phases and different input powers. We typically took data for 60 different phases for every input power within a range of a little less than $-\pi/2$ to $\pi/2$.

\subsection{Relation between detuning and quadrature}\label{ss:detuning_and_quadrature}

\begin{figure}[ht!]
\begin{center}
\includegraphics{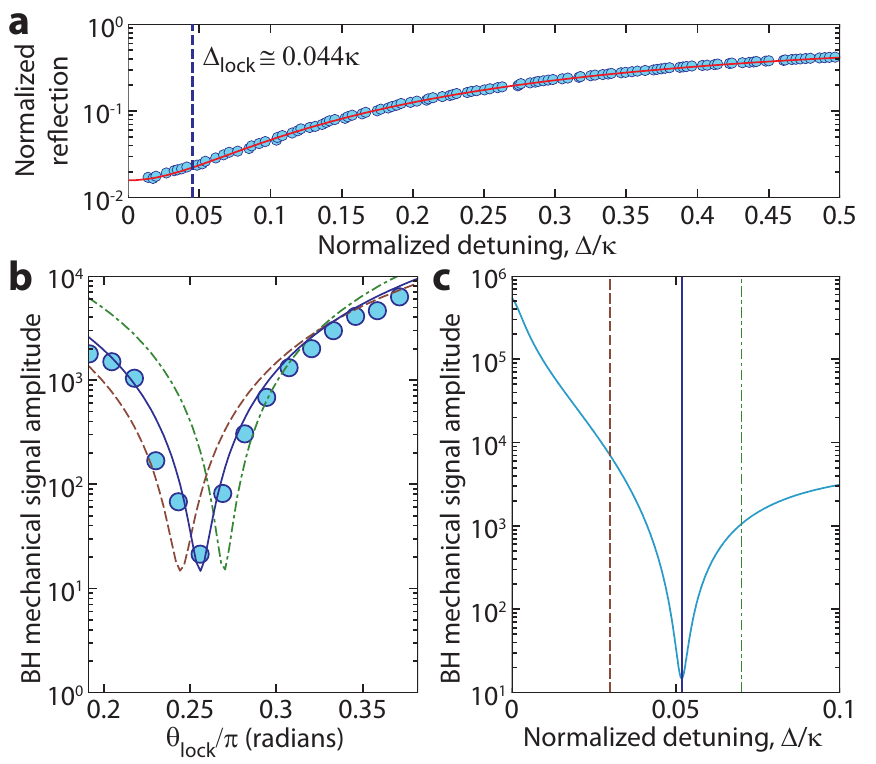}
\caption{\textbf{Detuning and phase lock points.} \textbf{a}, An optical scan taken  before the data run starts is shown. The blue vertical line denotes the target detuning the software lock moves the laser to, determined from the measured reflection intensity. The laser is kept at that detuning via  a wavemeter lock, as the light is switched away from PD1 and to the homodyne detector. The measured area under the mechanical mode is plotted in \textbf{b} (blue circles) at this detuning. A minimum value is reached for a local-oscillator to reflected signal phase of $\theta^\ast_\text{lock}$. Depending on the detuning, different mechanical mode amplitudes can be measured at this phase angle $\theta_\text{lock}$, according to the model. We obtain an accurate estimate of the detuning by calculating the detuning at which the mechanical mode amplitude is minimized at the measured $\theta_\text{lock}$ as shown in \textbf{c}. The expected mode amplitudes for the detunings represented by the red and green lines in \textbf{c} are shown by the similarly colored curves in \textbf{b}. }
\label{fig:noise_modelsX}
\end{center}
\end{figure}

The laser frequency is positioned at a detuning of roughly $0.04\cdot\kappa$ by starting at a larger detuning on the red side of the cavity, and stepping the laser blue in $0.1$~pm steps (12~MHz) towards the cavity while monitoring the average intensity of the reflected light on PD1. Once the target intensity is reached, the laser is kept at this wavelength during the course of the measurement by the wavemeter lock without further feedback from PD1. The intensity reading gives us an idea of the value of the detuning which is determined more accurately by analysis of our homodyne spectra. 

The homodyne spectra are taken at different phase lock points (see Fig.~\ref{fig:phases}) corresponding to quadrature angles $\theta_\text{lock}$ between the reflected signal and the LO. These angles differ from our convention in Section~\ref{sec:theory_squeezing} where the phase $\theta$ between the \textit{input} light into the cavity and the local oscillator is considered. They are related to oneanother by the phase imparted on the input light upon reflection from the cavity, 
\bea
\phi(\Delta) = \text{Arg} \left[ 1 - \frac{\kappa_e}{i\Delta + \kappa/2} \right],
\eea
and the relation
\bea
\theta_\text{lock} = \theta - \phi.
\eea
For a given laser-cavity detuning $\Delta$, we sweep through the different phase lock points  (see Fig.~\ref{fig:phases})  $\theta_\text{lock}$, and take mechanical spectra for each phase. The phase that minimizes the mechanical signal $\theta^\ast_\text{lock}$ is determined from the recorded spectra. This allows us to solve for $\Delta$ using the expression $\theta^\ast_\text{lock} = \theta^\ast(\Delta) - \phi(\Delta)$ where $\theta^\ast(\Delta)$ is the phase minimizing the mechanical transduction according to the model in the previous section. To first order (for $\Delta\ll \kappa$)  $\theta^\ast$ is  0 since no mechanical signal is observed in the intensity quadrature of the reflected light. This post-processing of the data allows us to determine that across the measured powers the detuning was $\Delta = 0.044\cdot\kappa \pm 0.006\cdot\kappa $. For a single measured power, we expect a more accurate determination, with an uncertainty on the order of  $0.003\cdot\kappa$. This level of accuracy in the detuning also determines the uncertainty in quadrature angle of 0.04 rad.

\section{Sample Fabrication and Characterization}

\subsection{Fabrication}
The devices are fabricated from a silicon on insulator (SOI) wafer (SOITEC, 220~nm device layer, $3~\mu\mathrm{m}$ buried oxide, device layer resistivity $4-20~\Omega\cdot\textrm{cm}$) using electron beam lithography and subsequent reactive ion etching (RIE/ICP) to form the structures. The buried oxide is then removed in hydrofluoric acid ($49\%$ aqueous HF solution) and the devices are cleaned in a piranha solution (3:1 H$_2$SO$_4$ and H$_2$O$_2$) and finally hydrogen terminated in diluted HF. For a more detailed description see~\cite{ChanPhD}.

\subsection{Optical Characterization}~\label{ss:optical_characterization}
The optical characterization of our devices is done by sweeping the laser frequency across the optical resonance while detecting the reflected light in a photodetector (PD1 in figure~\ref{fig:setup}). This light is simultaneously sent to a wavemeter to record the absolute wavelength and accurately determine the linewidth and center frequency of the resonance. Each chip contained several designs where the waveguide loading (coupling) of the optical cavity was varied by changing the gap size between the waveguide and nanobeam. For our measurements we chose a slightly overcoupled ($\kappa_\text{e}/\kappa_\text{i}\approx 1.22>1$) device with good optical quality (57,000 loaded $Q$)~\cite{Hill2013}.

\subsection{Mechanical Characterization}~\label{ss:thermometry}

\begin{figure*}[ht!]
\begin{center}
\includegraphics[width=2\columnwidth]{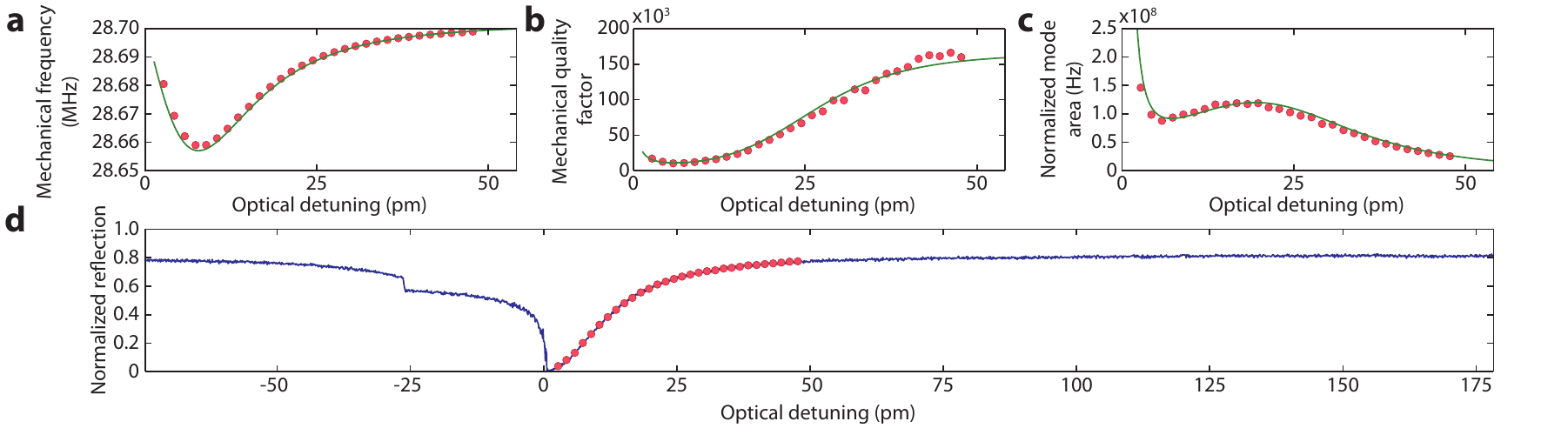}
\caption{\textbf{Optomechanical characterization.} We characterize the behavior of the optomechanical system in order to extract several parameters such as the intrinsic mechanical linewidth $\gamma_\text{i}$, the optomechanical coupling rate $g_0$, and the bath temperature $T_{\mathrm{b}}$ ($n_\text{b}$). \textbf{a}, The effective mechanical frequency $\omega_{\text{m}} = \omega_{\mathrm{m0}} + \delta\omega$ described in equation~\eqref{eq_spring} is plotted as a function of the laser detuning $\Delta = \omega_\text{o}-\omega_\text{L}$ (shown here in units of wavelength). The frequency shift is due to the optical spring effect caused by radiation pressure. \textbf{b}, The optomechanical interaction also causes the intrinsic linewidth $\gamma_\text{i}$ of the mechanical mode to be broadened as the detuning is changed (cf.\ equation~\eqref{eq_damp}). \textbf{c}, The area under the mechanical Lorentzian is also modified depending on $\Delta$, and is shown here, normalized to shot-noise. The fits (green lines) in \textbf{a}--\textbf{c} are now used to obtain $\gamma_\text{i}$, $g_0$ and $n_\text{b}$ (see text for details). The plot in \textbf{d} shows a normalized cavity scan, which is used to determine the exact detunings in \textbf{a}--\textbf{c}, with every red data point corresponding to a data point in \textbf{a}, \textbf{b} and \textbf{c}.}
\label{fig:thermometry}
\end{center}
\end{figure*}

The intrinsic mechanical damping rate $\gamma_\text{i}$ and the optomechanical coupling rate $g_0$ are measured by detecting the mechanical response to the signal laser, through the reflected signal field, on the spectrum analyzer. We keep the optical power constant, while we take measurements for several different detunings $\Delta$. The radiation pressure force causes both an optical spring effect resulting in a  frequency shift of the mechanical resonance, as well as damping of the mechanical motion, associated with a broadening of its linewidth (see equations~\eqref{eq_spring} and~\eqref{eq_damp}). By fitting the data shown in figures~\ref{fig:thermometry}a and~\ref{fig:thermometry}b, we can extract $\gamma_\text{i}=172$~Hz and $g_0=2\pi\times 750$~kHz. Knowing the mechanical properties of our resonator and the precise intracavity photon number, we can now also extract the mechanical bath occupancy $n_\text{b}$ as a function of detuning from the measured PSDs of the mechanical mode (figure~\ref{fig:thermometry}c; see also~\cite{Chan2011}). This shows us that our mechanical mode thermalizes to about 16~K for low optical input powers, which is close to the cold finger temperature of our cryostat of 10~K.

\section{Noise Spectroscopy Details}

\subsection{Homodyne measurement with laser noise}
Our experiment is designed to measure the spectral density of the fluctuations of the optical field exiting the cavity. However any real laser system will have technical noise, in addition to the quantum noise associated with an ideal coherent source, which adds to the detected noise level. Both the signal and local oscillator arm of our setup contain this noise which must be taken into account. The noise on the signal arm can also be modified non-trivially by propagation through the optomechanical system.  We start by reproducing known results on the operation of an ideal, balanced homodyne detection system with signal and local oscillator input fields $\op{a}{\mathrm{s}}$ and $\op{a}{\mathrm{LO}}$ respectively, under the influence of noise~\cite{Yuen1983,Schumaker1984,Shapiro1985}.
Most generally, these fields consist of coherent tones $\alpha_\mathrm{s}$ and $\alpha_\mathrm{LO}$, technical (or classical) noise components $a_{\mathrm{s,N}}(t)$ and $a_{\mathrm{LO,N}}(t)$, and quantum fluctuations $\op{a}{\mathrm{s,vac}}(t)$ and $\op{a}{\mathrm{LO,vac}}(t)$:
\bea
\op{a}{\mathrm{s}} &=& \alpha_\mathrm{s} + a_{\mathrm{s,N}}(t) + \op{a}{\mathrm{s,vac}}(t),\\
\op{a}{\mathrm{LO}} &=& \alpha_\mathrm{LO} + a_{\mathrm{LO,N}}(t) + \op{a}{\mathrm{LO,vac}}(t).
\eea
Since both the local oscillator field and the signal field are generated by the same laser, the technical noise on the signal and local oscillator will be correlated, and these correlations must be accounted for in the analysis. For the simplest case, where the signal arm does not experience the complex dispersion from interaction with an optomechanical system (e.g.\ being reflected off the end-mirror far detuned from the optical resonator), we expect
\bea
a_{\mathrm{s,N}}(t)  = \alpha_\mathrm{s} \xi(t)~~\text{and} ~~a_{\mathrm{LO,N}}(t)  = \alpha_\mathrm{LO} \xi(t).\label{eqn:noise_ideal_defn}
\eea
The function $\xi(t)$ is related to the intensity and phase fluctuations of the laser light ($n(t)$ and $\phi(t)$ respectively):
\bea
a(t) &=& a_0 (1+n(t)) e^{i\phi(t)} \approx a_0 (1+n(t) + i \phi(t))\nonumber\\
&&\xi(t) = n(t) + i\phi(t)
\eea

The difference of the photocurrent in the homodyne detector is given by 
\bea
\hat{I}(t) = \op{a}{\mathrm{s}} \opdagger{a}{\mathrm{LO}} + \opdagger{a}{\mathrm{s}} \op{a}{\mathrm{LO}},
\eea
which, considering only the technical noise, reduces to 
\bea
\hat{I}(t) = |\alpha_\text{LO}| \opX{X}{\text{s,vac}}{\theta} + I_\text{DC} ( 1 + 2 \text{Re}\{ \xi(t) \}), \label{eqn:bhd_noise}
\eea
under the assumption that $\alpha_\text{LO} \gg \alpha_\text{s}$, using the definitions in equation (\ref{eqn:noise_ideal_defn}), and taking the DC current $I_\text{DC} = 2 \text{Re}\{ \alpha_\text{s}^\ast \alpha_\text{LO} \} = 2 |\alpha_\text{s}\alpha_\text{LO}| \cos(\theta)$, where $\theta$ is the relative phase between the signal and local oscillator. From this equation we see that the phase noise $\phi(t)$ cannot be detected on a balanced homodyne setup. This can be understood as being from the detectors fundamental insensitivity to phase noise on the laser, as the only phase reference in the system is the local oscillator, which contains the same phase fluctuations as the signal. Secondly, for the local oscillator phase which makes $I_\text{DC} = 0$, intensity noise is not detected. In a real homodyne detector this is only true for a perfect common mode rejection ratio (CMRR), which is the case in our setup as the intensity noise is negligible and the CMRR is $>25$~dB. For these reasons we use a different setup for characterizing the laser phase and intensity noise as described in Section~\ref{ss:laser_phase_noise_measurement}.

\subsection{Measurement and characterization of laser noise}\label{ss:laser_phase_noise_measurement}

\begin{figure}[ht!]
\begin{center}
\includegraphics[width=1\columnwidth]{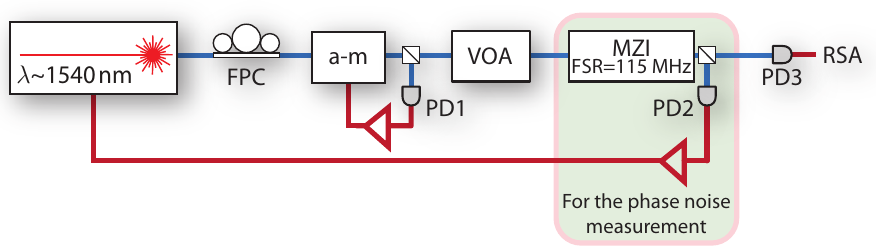}
\caption{\textbf{Experimental setup for characterization of intensity and phase noise.}  The laser is amplitude stabilized and an attenuator is used to select the desired optical power.  For the phase noise measurement the light is sent through a Mach-Zehnder interferometer (MZI) with a free spectral range of 115 MHz.  The laser is locked to the center of the interference fringe allowing frequency noise to be converted to intensity noise. The light is then detected on a New Focus Model 1811 photodetector and the photocurrent detected on a spectrum analyzer.  The same setup is used to detect intensity noise without the MZI.}
\label{fig:setup-phasenoise}
\end{center}
\end{figure}

\begin{figure*}[ht!]
\begin{center}
\includegraphics[width=2\columnwidth]{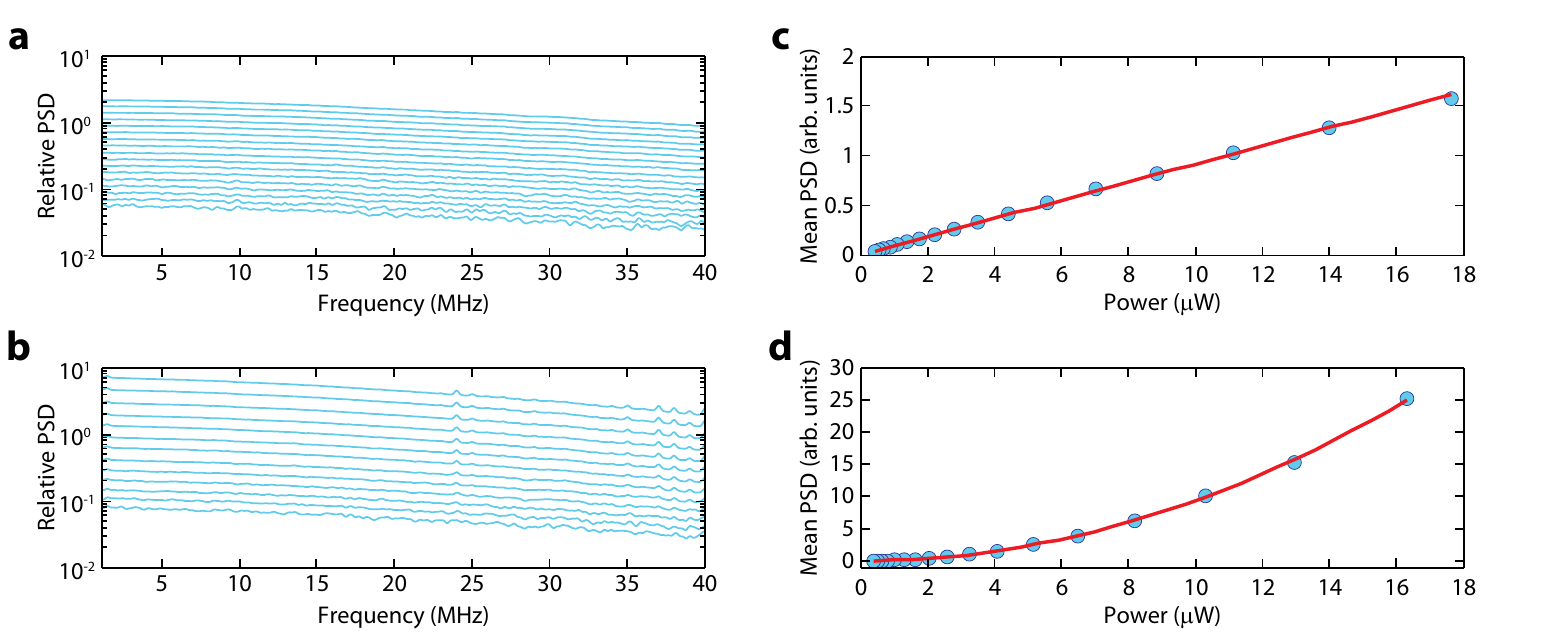}
\caption{\textbf{Laser noise characterization.} \textbf{a}, We measure the power spectral density (PSD) of our laser for several powers, normalize to and then subtract it from the dark current of the detector. The same measurement is performed using a Mach-Zehnder interferometer locked at half of the fringe amplitude in order to convert any frequency noise to intensity noise to allow detection and is shown in \textbf{b}. \textbf{c}, Plot of the mean value of the PSD around the mechanical frequency $\omega_{\mathrm{m}}$ from the measurement done in \textbf{a} as a function of power. The good linear fit (red line) indicates that no intensity noise is present. \textbf{d}, Mean PSD of the measurement in \textbf{b}. The quadratic fit (red line) shows that phase noise is indeed present (see text for more details).}
\label{fig:phase_noise}
\end{center}
\end{figure*}

In this section we discuss the procedure used for characterization of our laser (New Focus TLB-6728-P-D). This characterization was done using an independent setup, shown in figure~\ref{fig:setup-phasenoise}, and involved two measurements directly detecting the light.

The first measurement is to characterize the intensity noise where the laser light is sent directly onto a photodetector with the incident power varied.  From the theory we expect for the detector photocurrent
\bea
I(t) &=& (\alpha_\mathrm{LO} + a_{\mathrm{LO,N}}(t) + \op{a}{\mathrm{LO,vac}}(t))^\dagger ( \text{h.c.} ) \nonumber\\
 &=& |\alpha_\text{LO}| \opX{X}{\text{LO,vac}}{\theta=0} + I_\text{DC} ( 1 + 2 \text{Re}\{ \xi(t) \}),\nonumber\\
\text{with}&&I_\text{DC} = |\alpha_\text{LO}|^2.\nonumber
\eea
The spectral density of the current is then given by
\bea
S_{II}(\omega) = |\alpha_\text{LO}|^2 \left( 1 + |\alpha_\text{LO}|^2 S_{nn}(\omega) \right),
\eea
where $S_{nn}(\omega)$ is the PSD of the intensity noise fluctuations.
For a real detector, this equation is modified by the presence of a dark current $S_\text{dark}(\omega)$ and non-unity efficiency ($\eta_\text{det} <1$):
\bea
S_{II}(\omega) &=& S_\text{dark}(\omega)  \nonumber\\&& +|\alpha_\text{LO}|^2 \left( 1 + \eta_\text{det} |\alpha_\text{LO}|^2 S_{nn}(\omega) \right). \label{eq:intensity_noise}
\eea
We subtract the dark current (measured with the laser turned off) from the data, and set bounds on the magnitude of the intensity noise present in the laser by examining the linear and quadratic dependence of the noise floor with respect to power. The linear component is due to shot-noise, while the quadratic variance is due to the intensity noise fluctuations (see equation~\eqref{eq:intensity_noise}). The results are shown in figure~\ref{fig:phase_noise}a and c. The noise floor was seen to only increase linearly with laser power, confirming the absence of intensity noise at the frequencies of interest.

A second measurement is done to characterize the phase noise properties of the system. By sending the laser through a Mach-Zehnder interferometer (MZI) with transfer function $I(t) = I_0 (1 + \sin(2\pi\omega/\omega_\text{FSR}))$, the intensity of the transmitted light will contain fluctuations related to the frequency fluctuations of the light (see figure~\ref{fig:phase_noise}). The free spectral range (FSR) of the MZI is $\omega_{\mathrm{FSR}}/2\pi$ = 115~MHz. For a real detector, and assuming $\omega \ll \omega_\text{FSR}$, we arrive at
\bea
S_{II}(\omega) = S_\text{dark}(\omega) + |\alpha_\text{LO}|^2 \left( 1 + \eta_\text{det} \frac{|\alpha_\text{LO}|^2}{\omega^2_\text{FSR}} S_{\phi\phi}(\omega) \right).\nonumber
\eea
Some phase noise was detected, as shown in figure~\ref{fig:phase_noise}b and d and the quadratic dependence of the PSD on signal power. The spectral densities show a roll-off due to the FSR of the MZI. It was found that in the frequency range of interest, $1~\text{MHz} <\omega/2\pi < 40~\text{MHz}$, the frequency noise spectral density, $S_{\omega\omega}(\omega) = \omega^2 S_{\phi\phi}(\omega)$, is flat, and roughly equal to $3-6\times10^3~\text{rad}^2\cdot\text{Hz}$, in agreement with previous characterization of the same lasers at higher frequencies~\cite{Safavi-Naeini2013a}.

\subsection{Linearity of detector with local oscillator power}

\begin{figure}[ht!]
\begin{center}
\includegraphics[width=1\columnwidth]{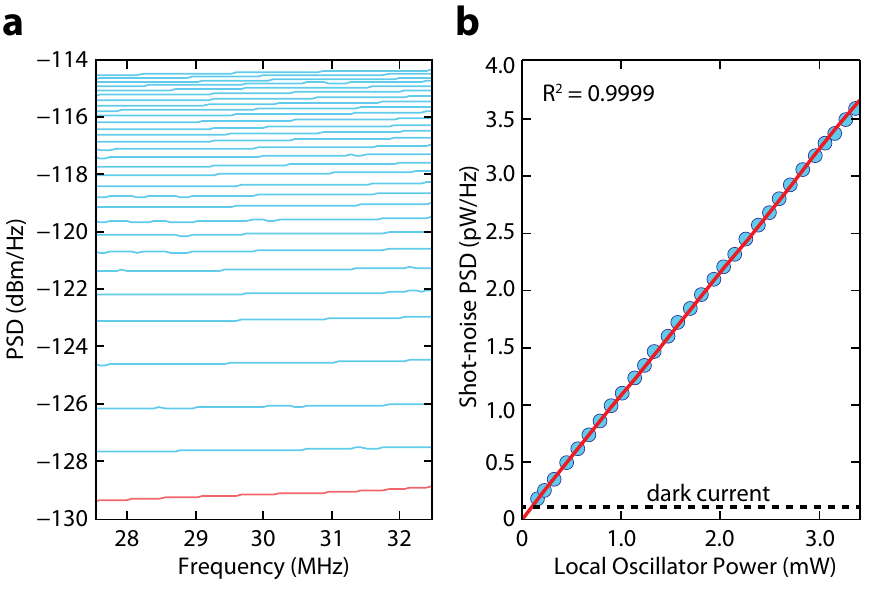}
\caption{\textbf{Noise level versus power.} \textbf{a}, Electronic noise power spectral densities from the balanced homodyne detector at different local oscillator powers (under a balanced condition). The red trace corresponds to the electronic noise floor with zero local oscillator power, i.e. the dark current. \textbf{b}, Mean value of the power spectral densities shown in \textbf{a} as a function of local oscillator power. In this plot the electronic noise or dark current contribution (0.12 pW/Hz, shown by the dashed black line) is subtracted. The red line is a linear fit, which has a coefficient of determination R$^2=0.9999$. The local oscillator power used in the experiment presented in the main text corresponds to 3.0 mW. }
\label{fig:snl_vs_lo_power}
\end{center}
\end{figure}

Having characterized the laser with an independent setup, we try to understand the properties of the measurement system. Our first measurement is designed to characterize the linearity of the detector and amplifier. With $I_\text{DC}=0$, and no signal in the signal arm of the BHD, we expect the system to faithfully reproduce the relation~(\ref{eqn:bhd_noise}) showing a linear relationship between local oscillator power and the detected signal vacuum fluctuation (shot-noise) noise level. It is observed that the mean value of the PSDs linearly depend on the input power as expected and shown in figure~\ref{fig:snl_vs_lo_power}. This indicates that our detector (and its amplifier) are in fact linear. The red line is a linear fit, with a coefficient of determination of R$^2=0.9999$. Although we already confirmed that no measurable amount of intensity noise is present (cf.\ figure~\ref{fig:phase_noise}), in the case we would have an appreciable amount of noise this measurement would show that it is smaller than the CMRR.

\subsection{Detected noise level with unbalancing}

\begin{figure}[ht!]
\begin{center}
\includegraphics[width=1\columnwidth]{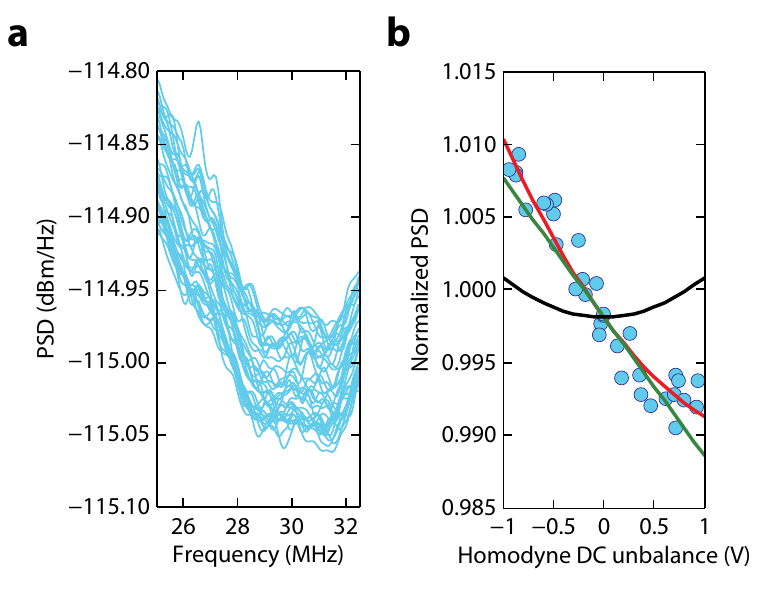}
\caption{\textbf{Amplifier gain.} \textbf{a}, Shown are the power spectral densities (PSDs) of the local oscillator as a function of the balancing of the optical power in the two paths of the homodyne detector. Each trace represents a different ratio of power in each path. These traces were taken with a local oscillator power of 3.0 mW, as used in the experiment. \textbf{b}, The mean value of the PSDs normalized to the perfectly balanced PSD are shown as a function of the difference voltage on the two photodiodes in the homodyne detector, where zero voltage represents perfect balancing. The green line is a linear fit to the data, while the black curve is a quadratic model, which describes any classical intensity noise that could cause the difference in the level of the PSDs. The red curve is the sum of the two. The change in PSD with homodyne unbalancing can be fully explained by the small signal gain weakly dependent on the detector unbalancing (linear fit) and no classical intensity noise (as previously determined).}
\label{fig:gain_vs_unbalance}
\end{center}
\end{figure}

A second measurement with vacuum input on the signal is done to understand how the amplifier in the homodyne detector depends on the DC level of the electronic signal after the photocurrent is subtracted. Here we use the variable coupler to change the splitting ratio and cause an imbalance between the optical power levels in the arms. The detected noise floors are shown in figure~\ref{fig:gain_vs_unbalance}a, and the mean detected PSDs are shown in figure~\ref{fig:gain_vs_unbalance}b, normalized to the shot-noise level. We find that at larger $V_\text{DC}$ (related to $I_\text{DC}$ linearly), there is a very small ($<2\%$) drop in the gain of the detector.  Using a linear fit, we extract an adjustment to the gain vs.\ output DC current. This means that for a measured noise power spectral density $S_\text{meas}(\omega)$ taken at a DC voltage $V_\text{DC}$, we estimate that the actual PSD, compensating for modified gain, is $S(\omega) = (1+V_\text{DC}/(-0.0096))^{-1} S_\text{meas}(\omega)$. This modification is used from here on, and only reduces the amount of squeezing we observe, as the quadratures with squeezing are always at positive voltages. Additionally, the largest DC voltages we work at are roughly $\pm 1$ V, which results in a modification on the order of one percent.

\begin{figure}[ht!]
\begin{center}
\includegraphics[width=1\columnwidth]{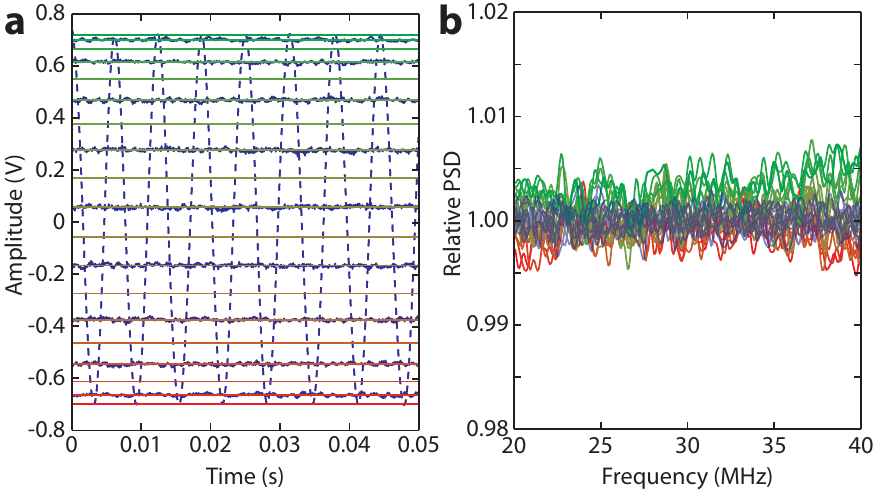}
\caption{\textbf{Detuned noise.} The laser is detuned with respect to the cavity resonance by 1~nm and spectra are taken using the homodyne detector over a range of phase angles, with a local oscillator power of 3.0~mW. This lets us estimate the amount of additional intensity noise we might acquire in our optical signal train. \textbf{a}, The dotted blue line shows the amplitude of the interference of the signal and local oscillator as a function of time. We lock at several relative phases (color-coded from green to red in \textbf{a} and \textbf{b}) and plot the associated normalized power spectral densities (PSD) relative to shot-noise in \textbf{b}. For every second measurement we switch the signal beam off to obtain the shot-noise level (blue traces). The maximum difference in the noise level is around 0.5\%.}
\label{fig:detuned_noise}
\end{center}
\end{figure}

\subsection{Estimating added noise in the optical train}

In our third measurement, we reflect the laser light off the end mirror of the waveguide coupler (detuned by $1$ nm from the cavity), and measure the detected noise level as a function of $\theta$, the phase difference between local oscillator and signal. This measurement is sensitive to both the conversion of phase noise to intensity noise through dispersion in the optical train, and added noise due to additional noise processes in the optical train such as guided acoustic-wave Brillouin scattering (GAWBS)~\cite{Shelby1985}, which could cause uncorrelated noise in the local oscillator and signal arms (see Eq.~\eqref{eqn:bhd_noise}). The results of this measurement are shown in figure~\ref{fig:detuned_noise}a and b. The first figure shows the DC interference signal between the local oscillator and signal used in the measurement. The LO power used for this experiment was the same as for the actual squeezing data, and the signal level used is on the same order as used for the highest power measurements, as is evident from the swing of about $1.4~\text{V}$ in the DC interference signal. The highest DC swing observed in the experiment was $1.6~\text{V}$. The second figure shows the normalized (to shot-noise) power spectral density where an increase of at most $0.5\%$ is observed, indicating these sources of noise do not contribute in our experiment.

\subsection{The effect of laser phase noise}

Using the measured value for the spectral density of phase and frequency fluctuations from section~\ref{ss:laser_phase_noise_measurement}, the effect of laser technical noise on the detected squeezing spectra can be calculated. Following the derivation in section~\ref{ss:detailed_squeezing_derivation} and taking the classical noise component of the field input to the cavity to be $a_\text{in}^{(\text{N})}(\omega) = i\alpha_\text{in} \phi(\omega)$ (with a corresponding LO phase noise of $a_\text{LO}^{(\text{N})}(\omega) = i\alpha_\text{LO} \phi(\omega)$), we arrive at an expression for the output noise due to input phase noise from the cavity:
\bea
a_\text{out}^{(\text{N})}(\omega)  = i \alpha_\text{in} (1 + A_1(\omega) - A_2(\omega)) \phi(\omega).
\eea
Without optomechanical interaction ($G = 0$) we find $A_1(\omega) = -\kappa/(i(\Delta-\omega)+\kappa/2)$, and $A_2(\omega) = 0$. We calculate the expression for the current noise due to laser phase noise using this expression: 
\bea
I^{(\text{N})}(\omega) &=& \alpha_\text{LO}^\ast a_\text{out}^{(\text{N})}(\omega) + \alpha_\text{LO} [a_\text{out}^{(\text{N})}(-\omega)]^\ast \nonumber\\&&
+\alpha_\text{out}^\ast a_\text{LO}^{(\text{N})}(\omega) + \alpha_\text{out} [a_\text{LO}^{(\text{N})}(-\omega)]^\ast\nonumber\\
&=& F(\omega) \phi(\omega)
\eea
with $F(\omega) = i|\alpha_\text{LO}\alpha_\text{in}| [e^{-i\theta} (r(\omega) - r(0))
   + e^{i\theta} (r(-\omega) - r(0))^\ast] $. The PSD of the photocurrent due to phase noise is found to be
\bea
S^{(\text{N})}_{II}(\omega) = |F(\omega)|^2 S_{\phi\phi}(\omega).
\eea
For a system with no dispersion, $r(\omega) = \text{const.}$, it can be easily shown that $F(\omega)=0$ as expected. For an over-coupled cavity with no optomechanical coupling, $r(\omega) = 1 - \kappa/(i(\Delta - \omega) + \kappa/2)$, so $r(\omega) - r(0) \approx 4i\omega/\kappa$, and we have $F(\omega) = 8  i|\alpha_\text{LO}\alpha_\text{in}|\sin(\theta)(\omega /\kappa)$. The $\omega$ dependence of $F(\omega)$ means that a flat frequency fluctuation spectrum ($S_{\phi\phi} \propto \omega^{-2}$, as we observe) adds a flat noise floor to the detected signal.

Finally we note that phase noise on the laser can drive the mechanical motion and cause heating. This effect is negligible since we are tuned near resonance, where only the intensity fluctuations affect the mechanics, and our cavity has a very large linewidth $\kappa$.

\subsection{Phenomenological dispersive noise model: the effect of structural damping} \label{ss:phenom_thermal_noise_model}

Mechanical damping of resonators and the associated fluctuations from coupling to the thermal bath has long been considered as an impediment to measuring weak forces in gravitational wave detectors~\cite{Saulson1990,Gillespie1993,Levin1998,Braginsky1999a,Liu2000}. In these studies the effect of the bath has often been encapsulated in a parameter $\Psi(\omega)$, representing the lag angle in the response of the material to a force. This lag angle is the complex part of the spring constant: $F = -k(1+i\Psi(\omega))x$.  The quality factor of the resonator is given by the narrow-band properties of the lag angle and its value at the mechanical resonance frequency, $Q=\Psi(\omega_\mathrm{m0})^{-1}$. We are interested in the wideband properties of $\Psi(\omega)$, since the spectral properties of the thermal fluctuations are related to the spectrum $\Psi(\omega)$, following the fluctuation-dissipation theorem. 

In the case of our experiments, we observed noise floors for $S_{xx}$ following a $\omega^{-1}$ power law on the low frequency end. This sort of noise power law corresponds to a flat spectrum for the lag angle $\Psi(\omega) = \text{const.}$ over the frequency range of interest. Unlike viscous damping which can be simply shown to have $\Psi(\omega) \propto \omega$ (since the force is proportional to velocity), a lag angle constant in frequency lacks a simple physical explanation, though it is ubiquitous in many types of mechanical resonators and commonly called ``structural damping''~\cite{Gillespie1993}.

In the input-output formalism outlined in section~\ref{sec:theory_squeezing} we model this type of noise by taking the mechanical damping rate $\gamma_\text{i}$ to be spectrally flat, and using  frequency dependent bath correlation functions
 $\avg{\op{b}{\mathrm{in}}(\omega)\opdagger{b}{\mathrm{in}}(\omega^\prime)} = (\bar{n}(\omega) + 1) \delta(\omega+\omega^\prime)$, $\avg{\opdagger{b}{\mathrm{in}}(\omega)\op{b}{\mathrm{in}}(\omega^\prime)} = \bar{n}(\omega) \delta(\omega+\omega^\prime)$, $\avg{\opdagger{b}{\mathrm{in}}(\omega)\opdagger{b}{\mathrm{in}}(\omega^\prime)} = 0$, and $\avg{\op{b}{\mathrm{in}}(\omega)\op{b}{\mathrm{in}}(\omega^\prime)} = 0$. This constitutes our single-mode thermal noise model.

\begin{figure}[ht!]
\begin{center}
\includegraphics{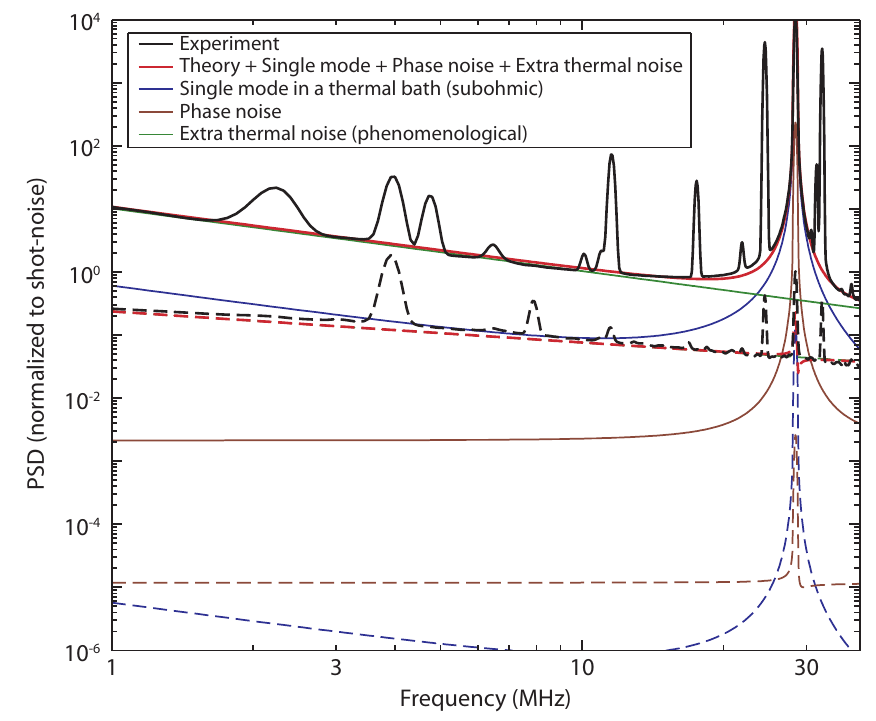}
\caption{\textbf{Noise model and experimental results.} The complete noise model, and constituent components, are plotted and compared to the experimental, shot-noise-subtracted power spectral density (PSD) for a quadrature sensitive to the mechanical motion (solid curves) and an insensitive quadrature (dashed curves).  The black lines are the experimental PSDs. The red lines represent the full noise model including contributions from a single mechanical mode (blue line), phase noise of the laser (brown line), and the extra thermal noise (green line) as described in section~\ref{ss:phenom_thermal_noise_model}. The deviation between the modeled and experimental data predominantly results from additional mechanical modes.}
\label{fig:noise_models}
\end{center}
\end{figure}

In any real optomechanical system, a family of mechanical modes couples to the optical resonance. In the modal picture which we use here, each of these mechanical resonances can be thought to add to the detected noise floor with its contribution scaling at the low-frequency end as $\omega^{-1}$. The contribution of each mode is proportional to the bath temperature, $g_\text{0,k}^2$,  $\gamma_\text{i,k}$, and $\omega_\text{m,k}^{-2}$. We lump all of these contributions into a single effective mechanical resonance, with its properties (not all independent) determined by fitting to the low frequency end of the noise floor. This mechanical resonator is modeled with a mechanical frequency $\omega_\text{m}/2\pi = 50~\text{MHz}$ (so we operate in the low frequency tail), a mechanical quality factor $Q_\text{m} = 100$, and a total coupling rate of $g_0/2\pi = 100~\text{kHz}$. We found that this model reproduced the magnitude and phase (the quadrature in which the noise is detected) of the $\omega^{-1}$ noise well, if an additional intracavity photon-dependent heating of $c_0 = 3.2\times 10^{-4}~$K/photons is assumed. These background noise floors are plotted in figure~\ref{fig:noise_models}. This cavity heating rate leads to the effective bath temperature to nearly double at the highest input powers, going from 16 K to over 30 K. This amount of heating is in line with what we expect from thin-film photonic crystals we have fabricated in the past operating in the same cryostat~\cite{Chan2011}.

\subsection{Phenomonological absorptive noise model} \label{ss:phenom_absorptive_noise_model}

\begin{figure}[ht!]
\begin{center}
\includegraphics{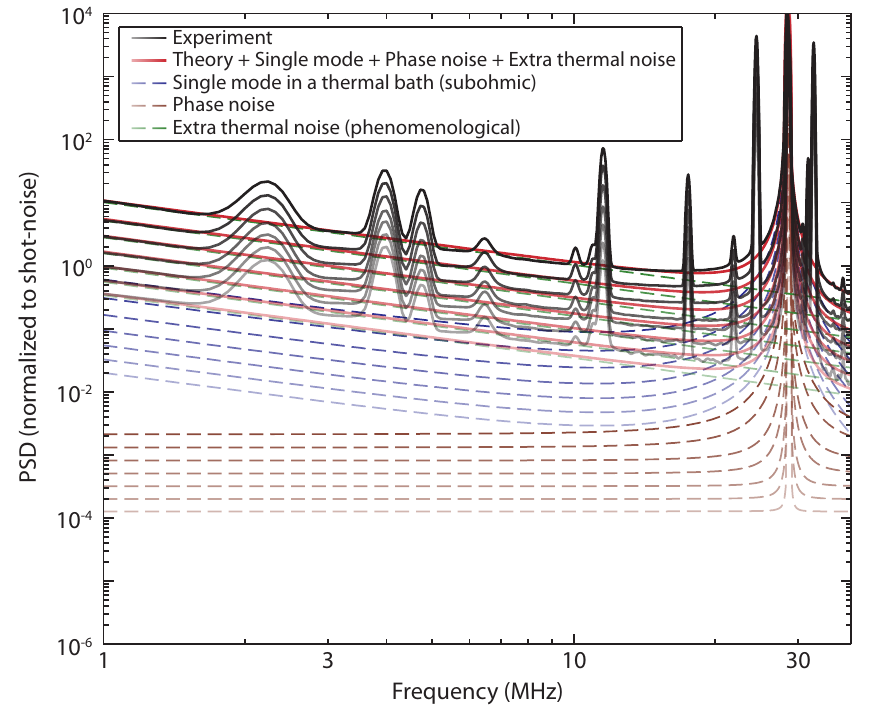}
\caption{\textbf{Power spectral densities (PSD) of noise contributions with varying powers.} The complete noise model along with its constituent components and experimental data are shown for varying optical powers. All curves have been normalized to the shot-noise level. The experimental data are shown in black with the full noise model in red consisting of the single mechanical mode (dashed blue), phase noise (dashed brown), and extra thermal noise (dashed green). The optical power scaling is represented by the transparency of the individual curves with curves becoming less transparent with increasing optical power.}
\label{fig:noise_vs_power}
\end{center}
\end{figure}

In addition to the noise in the quadrature of the mechanical motion (which arises from fluctuations in the cavity frequency $\omega_\text{o}$, and we suspect is mechanical in origin), we observed a significant amount of noise in the opposite quadrature, which can be interpreted to arise from fluctuations of the cavity decay rate $\kappa$. Additionally, we observed a different noise floor power law ($\omega^{-1/2}$) for this noise, which may rule out an optomechanical origin. The power law scaling agreed with thermorefractive noise studied extensively in the context of gravitational wave detection~\cite{Braginsky1999a},  microspheres~\cite{Gorodetsky2004}, and microtoroids~\cite{Schliesser2008a}, but it is expected that thermorefractive coupling is predominantly in the same quadrature as the mechanical noise, which is not observed here. Also, if thermorefractive, the noise should show strong variation with temperature through both a quadratic temperature scaling ($T^2$) and an extremely steep variation of $dn/dT$ in the temperature range of 16~K to 30~K~\cite{Komma2012}, which was not observed. At this point, we have no noise model to explain the observed fluctuations, and the origin of this noise will be the subject of further investigation to be presented at a later time. A phenomenological noise model was instead used, where fluctuations in the cavity linewidth proportional to the intracavity power with a $\omega^{-1/2}$ noise spectrum are assumed.

\begin{figure*}[ht!]
\begin{center}
\includegraphics{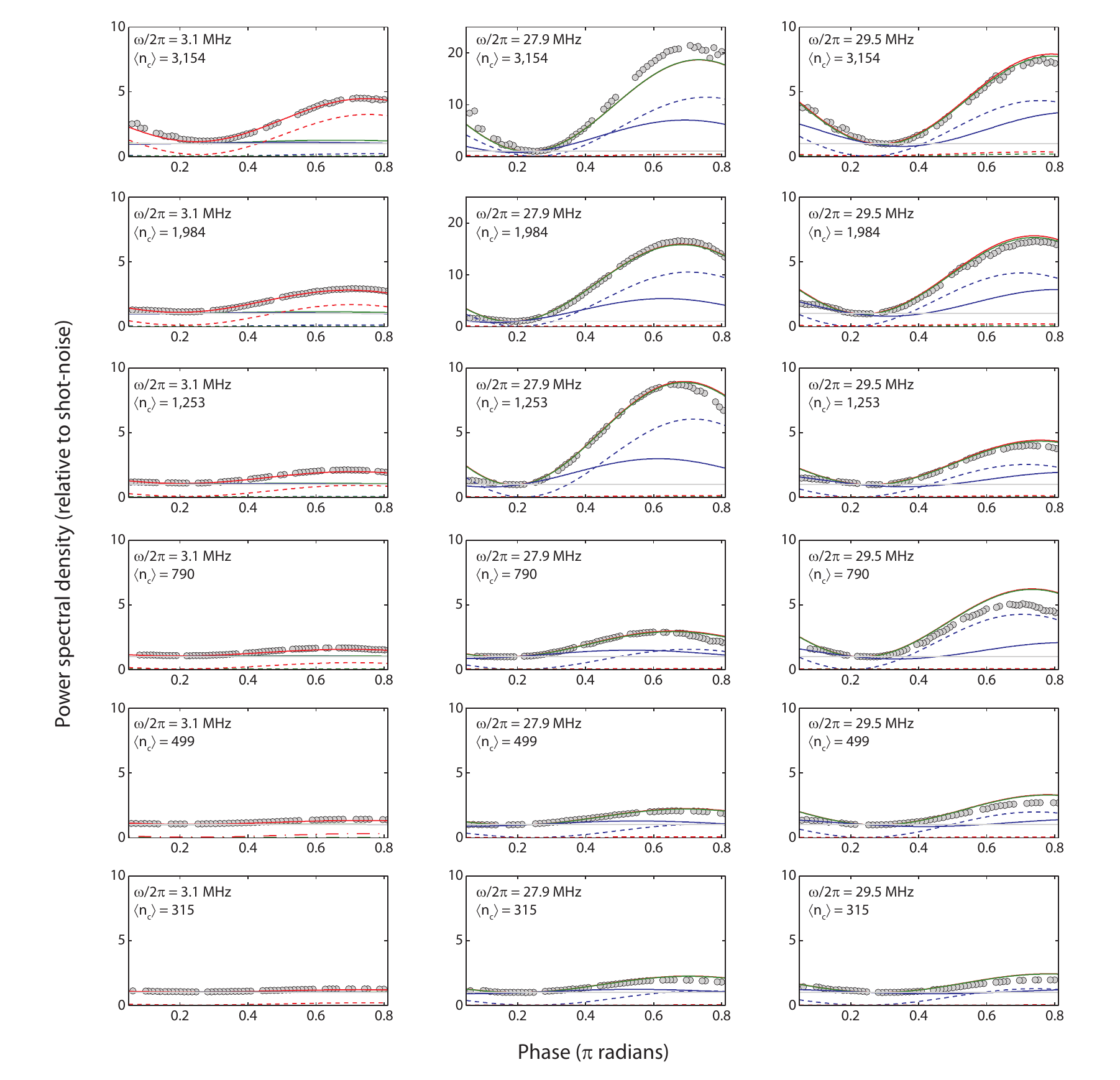}
\caption{\textbf{Detected noise power at a given frequency vs.\ the lock angle.} In these plots, a series of traces is shown of the detected noise level at a given frequency (with resolution bandwidth of 300 kHz) as a function of the locked phase $\theta_\text{lock}$. The grey points are the measured data points. The solid lines are the results of models detailed in this text, and the dashed lines represent different components of noise present in each model. The red line shows the full noise model, containing the transduced thermal brownian motion from the studied mode, the noise due to structural damping present in the system, the phase noise, and the phenomological out-of-quadrature noise. The green line is for a model considering all the same noise contributions, except the phenomological component. A model considering a system without any thermal noise is shown in blue. With no thermal force on the mechanical systems, the detected signal in this case can be attributed to radiation pressure shot-noise heating. The shot-noise level is denoted by a grey line. The contribution due to thermal motion of the mode of interest is shown by the dashed blue line. The noise contributions due to phase noise and structural damping are much smaller and shown by the green and red dashed lines, respectively.}
\label{fig:tomography_nzi}
\end{center}
\end{figure*}

\begin{figure*}[ht!]
\begin{center}
\includegraphics{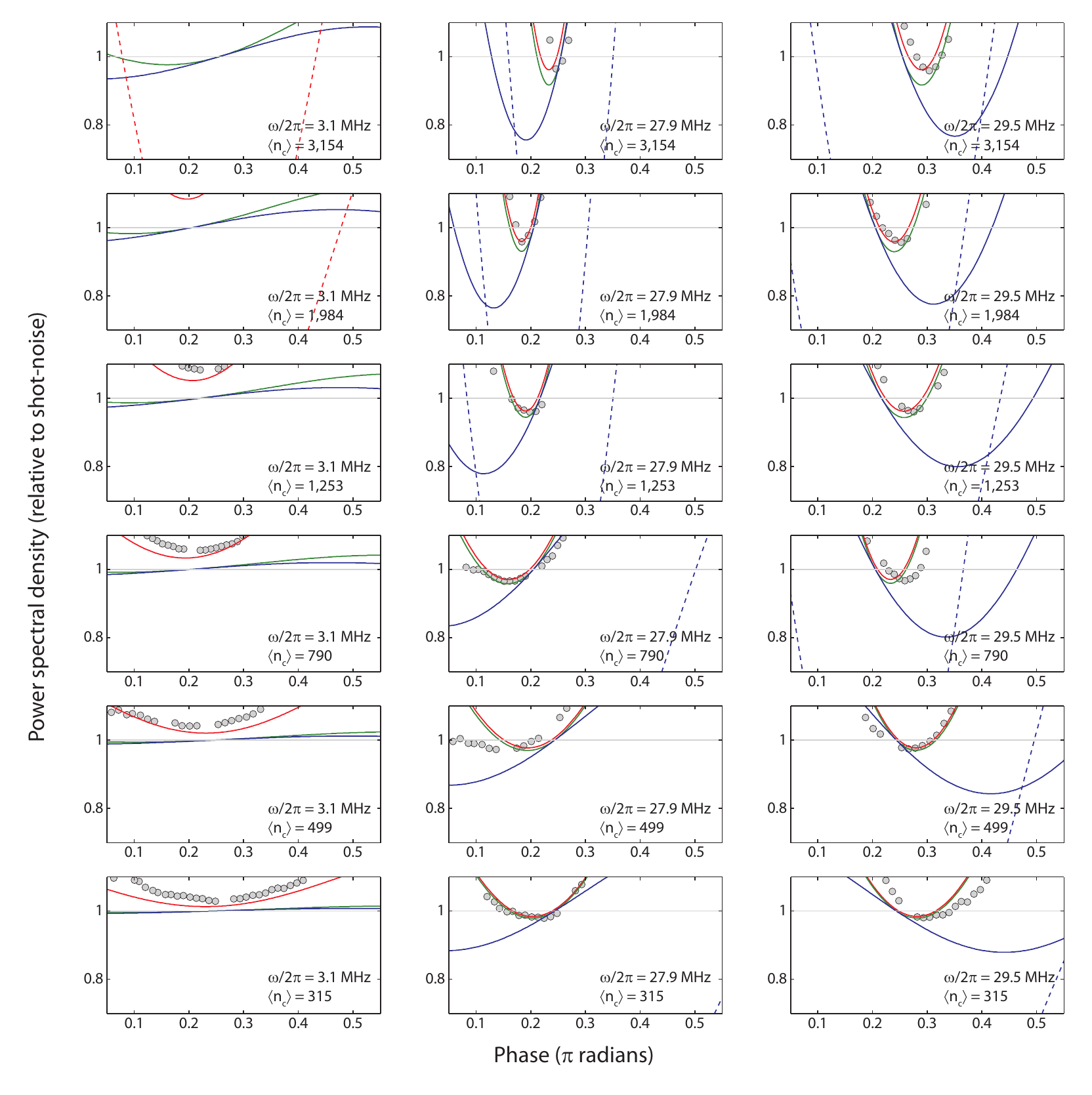}
\caption{\textbf{Close-up of detected noise power at a given frequency vs.\ the lock angle.} This close-up shows regions of squeezing, and the colors are the same as in Figure~\ref{fig:tomography_nzi}.}
\label{fig:tomography_zi}
\end{center}
\end{figure*}

\subsection{Comparing measured spectra to theoretically predicted spectra} 

Our spectrum analyzer (Tektronix RSA3408B) operates by taking Fourier transforms of a time domain signal. By windowing a short time sample, and calculating its energy spectrum, a power spectral density is constructed. The size of the window in the time-domain affects the resolution bandwidth, and is well known in signal processing, multiplication by a Gaussian window of length $\tau$ is equivalent to convolution of the frequency domain signal by a Gaussian with width proportional to $\tau^{-1}$. All of the measured data, except that presented in section~\ref{ss:thermometry} was taken with a 40~MHz window and 300~kHz resolution bandwidth. The spectra contain 501 points spaced by 80~kHz in the frequency domain. Additionally, for a few data sets we took narrow band spectra (down to 100~Hz resolution bandwidth) and found that the results agreed over the regions where squeezing was observed. The theory was calculated at 100 times finer resolution than the sampled data (with 50,000 points), and was then down-sampled after a Gaussian convolution step simulating the operation of the spectrum analyzer. This only effects the size of the mechanical peak, and has no effect on the frequency ranges where we see sub-shot-noise fluctuation spectra. For the thermometry data in section~\ref{ss:thermometry}, since we are interested in the mechanical linewidths and areas, the span was always chosen to be the minimum allowable by the RSA, which is twice as large as the linewidth.

\section{Summary of Noise Model}

In Table~\ref{tab:parameters} we present a summary of the parameters used in the theoretical model for the wideband squeezing spectra shown in the main text.

\renewcommand{\arraystretch}{1.1}
\renewcommand{\extrarowheight}{0pt}
\begin{table*}
\caption{Model Parameters}
\label{tab:parameters}
\begin{center}
\begin{tabularx}{\linewidth}{XlYX}
\hline
\hline
Symbol & Name & Value & Measurement \\
\hline
$Q_\text{optical}$ & Optical quality factor &  $5.7\times 10^4$ & Low-power optical spectroscopy with wavemeter.~\ref{ss:optical_characterization}\\
\hline
$\eta_\kappa$ & Cavity-waveguide coupling efficiency &  0.55 & Low-power optical spectroscopy with wavemeter. Verified phase response with VNA to distinguish from under-coupling.~\ref{ss:optical_characterization}\\
\hline
$\gamma_\text{i}/2\pi$ & Mechanical linewidth & 172~Hz & Linewidth measurement vs.\ laser detuning in thermometry measurement (see section~\ref{ss:thermometry}).\\
\hline
$g_0/2\pi$  & Optomechanical coupling rate & 750~kHz & Linewidth and mechanical frequency measurement vs.\ laser detuning in thermometry measurement (see section~\ref{ss:thermometry}).\\
\hline
$T^{0}_b$ & Bath temperature & 16~K & Calibrated areas in thermometry measurement (see section~\ref{ss:thermometry}).\\
\hline
$c_0$ & Heating by optical absorption & $3.2\times 10^{-4}$ K/photon & Rise of $\omega^{-1}$ noise floor with optical power (see section~\ref{ss:phenom_thermal_noise_model}). The cavity temperature according to this model rises from 16~K to roughly 30~K at the highest powers.\\
\hline
$S_{\omega\omega}$ & Frequency noise spectral density & $6\times 10^3~\text{rad}^2\text{Hz}$ & Frequency noise measurement with Mach-Zehnder Interferometer (see section~\ref{ss:laser_phase_noise_measurement}).\\
\hline
$\Delta$ & Laser detuning (red laser is positive) & $(0.044\pm0.006)\kappa$ & The intensity of the reflected light is used to initially set the detuning. For a more accurate determination, the value of $\Delta$ minimizing the detected signal for the observed $\theta^\ast_\text{lock}$ is found (see Section~\ref{ss:detuning_and_quadrature}).\\
\hline
$\theta_\text{lock}$ & lock angle & varies & The lock point (as in figure~\ref{fig:phases}) is used to find the phase angle between the light reflected from the cavity and the local oscillator.\\
\hline
$\theta^\ast_\text{lock}$ & critical lock angle & varies & This is the lock angle were no mechanical signal is detected. It is found by looking at the area of the mechanical mode as a function of $\theta_\text{lock}$ (see Section~\ref{ss:detuning_and_quadrature}).\\
\hline
\hline
\end{tabularx}
\end{center}
\end{table*}
\renewcommand{\arraystretch}{1.0}
\renewcommand{\extrarowheight}{0pt}

\section{Mathematical Definitions}
\label{sec:math_defn}

We present here the notational conventions used throughout this work for reference. The Fourier and inverse Fourier transforms of operator $\op{A}{}(t)$ are defined as
\bea
\op{A}{}(t) &=& \frac{1}{\sqrt{2\pi}} \int_{-\infty}^{\infty} d\omega~ e^{-i\omega t}\op{A}{}(\omega)~~~\text{and} \nonumber\\
\op{A}{}(\omega) &=& \frac{1}{\sqrt{2\pi}} \int_{-\infty}^{\infty} dt~ e^{i\omega t}\op{A}{}(t),
\eea
respectively. The Hermitian conjugate of operator $\op{A}{}(t)$ is given by $\opdagger{A}{}(t)$ which has the Fourier transform $\opdagger{A}{}(\omega)$. This is related to $\op{A}{}(\omega)$
\bea
\left(\op{A}{}(\omega)\right)^\dagger = \opdagger{A}{}(-\omega).
\eea
In the derivations presented here, we typically express a given operator in terms of the ``input'' bath operators. Expected values are then defined as $\avg{\op{A}{}(t)} = \text{Tr}[\rho_\text{in} \op{A}{}(t) ]$, where $\rho_\text{in}$ is the mixed state describing the bath. Spectral densities are found by taking the Fourier transform of the auto-correlation
\bea
S_{AA}(\omega) = \int_{-\infty}^{\infty} d\tau~ e^{i\omega \tau} \avg{\opdagger{A}{}(t+\tau)\op{A}{}(t)}.
\eea
For a real operator, $S_{AA}(\omega) = S_{AA}(-\omega)$. For non-real operators, the spectral density can be symmetrized to $\bar{S}_{AA}(\omega)$:
\bea
\bar{S}_{AA}(\omega) = \frac{1}{2} (S_{AA}(\omega) + S_{AA}(-\omega)).
\eea
These spectral densities can also be calculated from the frequency domain operators:
\begin{eqnarray}
S_{AA}(\omega) = \int_{-\infty}^{\infty} d\omega^\prime~ \avg{\opdagger{A}{}(\omega)\op{A}{}(\omega^\prime)}\label{eqn:SAA_omega_integral}.
\end{eqnarray}
We use the convention
\bea
\opX{X}{j}{\theta}= \op{a}{j} e^{-i \theta}+ \opdagger{a}{j}e^{i \theta}.
\eea
to define a measured quadrature of the field. This definition corresponds to having a phase difference of $\theta$ between input light and local oscillator.

\end{document}